\documentclass[aip,preprint,amsmath,amssymb,floatfix,jcp]{revtex4-2}

\usepackage{graphicx}
\graphicspath{{figures/}}
\usepackage{physics}
\usepackage{multirow}
\usepackage{xcolor}
\usepackage{caption}
\usepackage{subcaption}
\usepackage{xcolor}
\usepackage{bm}
\usepackage{eufrak} 
\usepackage{yfonts} 
\usepackage[title,titletoc]{appendix}
\usepackage{diagbox}
\usepackage{booktabs} 
\usepackage{algorithm}
\usepackage{algpseudocode}
\usepackage{amsfonts} 
\usepackage{dcolumn} 
\usepackage{siunitx} 

\usepackage{hyperref} 
\hypersetup{colorlinks=true, citecolor=blue, urlcolor=blue, linkcolor=blue}
\usepackage{cleveref}
  	\crefname{figure}{Figure}{Figures}
  	\crefname{table}{Table}{Tables}
  	\crefname{equation}{Eq.}{Eqs.}
  	\crefname{section}{Section}{Sections}
  	\crefname{subsection}{Section}{Sections}
  	\crefname{subsubsection}{Section}{Sections}
  	\crefname{algorithm}{Algorithm}{Algorithms}
  	\crefname{appendix}{Appendix}{Appendices}

\newcommand\vartextvisiblespace[1][.5em]{%
  \makebox[#1]{%
    \kern.07em
    \vrule height.3ex
    \hrulefill
    \vrule height.3ex
    \kern.07em
  }
}
\usepackage{todonotes}

\begin{document}

\title{Orbital Anatomy of Self-Interaction in Kohn-Sham Density Functional Theory} 

\author{Samuel A. Slattery}
\author{Edward F. Valeev}
\email{efv@vt.edu}
\affiliation{Department of Chemistry, Virginia Tech, Blacksburg, VA 24061}


\begin{abstract}
Self-interaction is a fundamental flaw of practical Kohn-Sham Density Functional Theory (KS DFT) approximations responsible for numerous qualitative and even catastrophic shortcomings. Whereas self-interaction is easy to characterize in one-electron systems, its orbital-dependent nature makes it difficult to uniquely define and decompose it and other components of the KS energy into orbital components. By starting from the orthogonal Hartree ansatz as the exchange-free self-interaction-free model, it is possible to uniquely define self-interaction and genuine exchange total energies as well as their decomposition into orbital contributions. Using Edmiston-Ruedenberg orbitals as a close numerical surrogate of the Hartree orbitals we investigate the accuracy of genuine exchange energies predicted by several levels of Jacob's ladder of functionals and their Perdew-Zunger (PZ) self-interaction corrected counterparts. Among several notable insights provided by the per-orbital breakdown of the KS energies the key is the delicate cancellation of errors between core and valence orbitals that belies the remarkable accuracy of many functionals (particularly, GGAs). Disruption of the cancellation by the PZ correction accounts for its relatively modest impact on GGAs and other semilocal functionals.
\end{abstract}

\maketitle

\section{Introduction}

The Kohn-Sham (KS) Density Functional Theory (DFT) accounts for the vast majority of electronic structure simulations of molecules and materials, due to its unrivaled accuracy/cost ratio. Although for ground states characterized by strong/nondynamical correlation (i.e., for states poorly approximated by a single Slater determinant) its accuracy is generally poor and it is often necessary to go to the more expensive many-body methods (such as coupled-cluster\cite{VRG:coester:1958:NP,VRG:shavitt:2009:} GW,\cite{VRG:hedin:1965:PR,VRG:aryasetiawan:1998:RPP} quantum Monte-Carlo,\cite{VRG:austin:2012:CR} or tensor-network methods\cite{VRG:white:1992:PRL,VRG:wouters:2014:EPJD}). However, even for states that are well-approximated by a single determinant KS DFT can produce large errors or even fail catastrophically.\cite{VRG:cohen:2008:S} Such errors are commonly attributed to the unphysical {\em self-interaction} (SI) of electrons\cite{VRG:fermi:1934:MAd,VRG:perdew:1981:PRB,VRG:mori-sanchez:2006:JCP} in approximate KS DFT which makes it not exact even for 1-e systems; the electronic Hamiltonian of course is self-interaction-free.

For many-electron systems it is harder to attribute the unphysical behavior of KS functionals to the self-interaction effects due to the {\em lack of (unique) definition of SI} in a many-electron system. Terms like ``many-electron self-interaction'' have been discussed\cite{VRG:mori-sanchez:2006:JCP} but other more generic and measurable terms like {\em delocalization error}\cite{VRG:johnson:2008:JCP, VRG:hait:2018:JPCL} are also used. Without a concrete definition of self-interaction it is not possible to measure it quantitatively, and one can at best discuss the {\em symptoms} of SI rather than how to eliminate its effects. Thus the first and foremost goal of this work is to try to define the self-interaction contribution to the KS energy as uniquely as possible.

Despite the lack of a unique definition, the self-interation is generally believed to cause of many shortcomings of practical KS approximations in many-electron systems and a key to improving the accuracy of KS DFT.\cite{VRG:perdew:2015:AIAMaOP}
Amelioration of self-interaction in self-consistent KS DFT has been attempted by (a) {\it a priori} modification of the KS formalism itself via various self-interaction corrections,\cite{VRG:tsuneda:2014:JCP} such as Perdew-Zunger (PZ) self-interaction correction\cite{VRG:perdew:1981:PRB} and its refinements\cite{VRG:zope:2019:JCP}), (b) {\it a posteriori} corrections of the KS DFT energy\cite{VRG:li:2018:NSR,VRG:mahler:2022:PRB}. Despite the fundamental differences in these routes, their designs are related by the common fact that the self-interaction is an orbital-dependent phenomenon. This is most trivially illustrated for the Coulomb energy of self-repulsion of the total charge density $n$ (often misleadingly referred to as the Hartree energy;\cite{VRG:engel:2011:} see \cref{section:formalism}):
\begin{align}
\label{eq:J}
    J = & \frac{1}{2} \int \frac{n(\mathbf{r})n(\mathbf{r'})}{\vert\mathbf{r}-\mathbf{r}'\vert} \, \mathrm{d}\mathbf{r} \, \mathrm{d}\mathbf{r}'
 \end{align}
Decomposing the charge density into its contributions $n_i$ from each KS orbital,
\begin{align}
\label{eq:ni}
n({\bf r}) = \sum_i n_i ({\bf r}) = \sum_i |\phi_i({\bf r})|^2
\end{align}
allows identification of the self-interaction contribution to $J$ as
\begin{align}
\label{eq:JSI}
J^\mathrm{SI} = \sum_i \frac{1}{2} \int \frac{n_i(\mathbf{r})n_i(\mathbf{r'})}{\vert\mathbf{r}-\mathbf{r}'\vert} \, \mathrm{d}\mathbf{r} \, \mathrm{d}\mathbf{r}' \equiv \sum_i [J^\mathrm{SI}]_i,
\end{align}
where $[\dots]_i$ will denotes a contribution from orbital $i$.
The charge density and the Coulomb energy, as quadratic forms of the orbitals, are invariant with respect to a unitary rotation of KS orbitals; in other words, they depend on the subspace spanned by the orbitals, but not the particular choice of the basis in that subspace. Thus there is no unique choice of orbital basis leading to a unique orbital decomposition of change density or energy.\cite{VRG:dutoi:2006:CPL}
In contrast, $J^\mathrm{SI}$ is a {\em quartic} form of the orbitals and is not unitarily invariant; changing orbitals will change orbital contributions $[J^\mathrm{SI}]_i$ as well as the total $J^\mathrm{SI}$. Hence to analyze the errors induced by self-interaction we must define the unique
orbital basis appropriate for such analysis. Note that the idea of orbital analysis of SI errors has been considered in, e.g., the context of Perdew-Zunger corrected KS DFT\cite{VRG:romero:2021:PCCP,VRG:vydrov:2004:JCP} and much earlier for the X$\alpha$ method;\cite{VRG:lindgren:1972:PRA} we are not aware of any such analysis in a more general context. The second factor that must be considered is the role of self-consistency. The self-interaction {\em directly} affects the energy and the effective energy operator (the Fock operator) which then introduce {\em indirect} effects through the self-consistent optimization of the orbital subspace. The distinction between direct and indirect effects of self-interaction has been drawn recently, for example, in the analysis of DFT errors in reaction barrier heights.\cite{VRG:kanungo:2024:JPCL,VRG:kaplan:2023:JCTC} Abandoning the self-consistency and using model SI-free orbitals has also been considered by Burke and co-workers.\cite{VRG:song:2022:JCTC,VRG:sim:2022:JACS}

In this work we reexamine the two issues that complicate the analysis of SI errors by focusing on the SI effects on the (mean-field) {\em exchange} energy (rather than combining the effects of SI on exchange and correlation together). We start by defining the {\em genuine} mean-field exchange as the difference between the Hartree-Fock and (orthogonal) Hartree energies, the latter being {\em the} SI-free and exchange-free mean-field theory. This in turn provides a concrete definition of the SI contribution to the exchange energy in a many-electron system. We show that the genuine exchange energy can be accurately and more robustly obtained from localized Hartree-Fock orbitals, with the best choice being the Edmiston-Ruedenberg orbitals\cite{VRG:edmiston:1963:RMP} due to their formal connection to the orthogonal Hartree method. Using such ER-HF-based genuine exchange as the reference we can thus analyze the functional-induced SI errors in exchange energies for a range of DFT functionals spanning the local density approximation (LDA), generalized gradient approximation (GGA), meta-GGA, and hybrid families. In addition to quantifying the SI errors in different functionals, the revealed orbital anatomy of SI errors shows systematic error cancellation between core and valence orbitals, both in atoms and molecules. The analysis also allows us to illustrate that the relatively smaller effect of PZ correction when combined with post-LDA functionals is due to spoiling the cancellation of SI errors. This can help us understand why the improvements from the PZ correction are uneven or even nonexistent for many molecular properties.\cite{VRG:perdew:2015:AIAMaOP}

The rest of this paper is structured as follows.
In \cref{section:formalism} we detail our formalism for orbital analysis of self-interaction errors in KS DFT.
Technical details of our computational experiments are described in \cref{section:tech-det-orb-anatomy}, and the findings are discussed in \cref{section:results-orb-anatomy}.

\section{Formalism}
\label{section:formalism}

\subsection{Self-Interaction in Mean-Field Models}
\label{subsection:self-interaction-mean-field-models}

The total electronic KS DFT energy,
\begin{align}
\label{eq:EKS}
 E^{\rm KS} = \sum_i \bra{i}\hat{h}\ket{i} + J + E^{\rm KS}_{\rm xc}
\end{align}
includes the 1-electron kinetic and external potential contribution $\hat{h}$, the Coulomb energy $J$ (\cref{eq:J}), and the exchange-correlation energy $E^{\rm KS}_{\rm xc}$. The latter is a functional of the total charge density $n$ alone in LDA and GGA functionals, as well as the orbitals $\{\phi_i\}$ (which are also formal functionals of $n$) in meta-GGA and generalized Kohn-Sham treatment of hybrid functionals. 
$E^{\rm KS}_{\rm xc}$
can be rewritten in terms of the exchange-correlation potential $v_\mathrm{xc}^\mathrm{KS}[n]$,
\begin{align}
  v_\mathrm{xc}^\mathrm{KS}[n](\mathbf{r}) = \frac{\delta E^{\rm KS}_{\rm xc}}{\delta n(\mathbf{r})},
\end{align}
as
\begin{align}
\label{eq:Exc}
E_{\mathrm{xc}}^{\mathrm{KS}} = \int v_\mathrm{xc}^\mathrm{KS}[n] (\mathbf{r}) \, n(\mathbf{r}) \, \mathrm{d}\mathbf{r}.
\end{align}
Combining \cref{eq:Exc} with \cref{eq:ni} allows one to decompose the exchange-correlation energy into orbital components:
\begin{align}
\label{eq:Exc_by_orb}
E_{\mathrm{xc}}^{\mathrm{KS}} = \sum_i \int v_\mathrm{xc}^\mathrm{KS}[n](\mathbf{r}) \, n_i(\mathbf{r}) \, \mathrm{d}\mathbf{r}.
\end{align}
This reveals that $E_{\mathrm{xc}}^{\mathrm{KS}}$ can also include the self-interaction for each electron since in general potential $v_\mathrm{xc}^\mathrm{KS}[n](\mathbf{r})$ depends on the state of each electron in the system.

For the KS model to become exact the self-interactions in $E_{\mathrm{xc}}^{\mathrm{KS}}$ must cancel exactly the self-repulsions in $J$ and it must be self-interaction-free otherwise. In practice, the density functional approximations do not ensure such cancellation. To better understand the anatomy of self-interaction let's first consider how the self-interaction is avoided in the Hartree-Fock method. The HF energy,
\begin{align}
\label{eq:EHF}
E^\mathrm{HF} = \sum_i \bra{i}\hat{h}\ket{i} + J + E_\mathrm{x}^\mathrm{HF},
\end{align}
is obtained from the KS energy (\cref{eq:EKS}) by replacing $E_\mathrm{xc}^\mathrm{KS}$
with
\begin{align}
\label{eq:ExHF}
    E_\mathrm{x}^\mathrm{HF} = & - \frac{1}{2} \sum_{ij} \braket{ji}{ij}
\end{align}
where
\begin{align}
  \braket{ij}{kl} = \int \int \frac{\phi_i^*(\mathbf{r}) \phi_k(\mathbf{r}) \phi_j^*(\mathbf{r}') \phi_l(\mathbf{r}')}{\vert\mathbf{r}-\mathbf{r}'\vert} \, \mathrm{d}\mathbf{r} \, \mathrm{d}\mathbf{r}'
\end{align}
are the usual Coulomb (2-electron) integrals over orbitals $\phi_p(\mathbf{r})$.
By rewriting \cref{eq:J,eq:JSI} in terms of orbitals also:
\begin{align}
\label{eq:J-phi}
    J = & \frac{1}{2} \sum_{ij} \braket{ij}{ij},\\
\label{eq:JSI-phi}
    J^\mathrm{SI} = & \sum_{i} \frac{1}{2} \braket{ii}{ii},
\end{align}
it is clear that the $J^\mathrm{SI}$ contribution occurs in $J$ and $E_\mathrm{x}^\mathrm{HF}$ with opposite signs, hence it cancels in the two-electron HF energy $J + E_\mathrm{x}^\mathrm{HF}$.

In the context of KS DFT there is no formal separation between exchange and correlation effects, however in practice the exchange and correlation effects are modeled separately, and the total exchange-correlation functional is a sum of the respective contributions. This makes sense since the physics of exchange and correlation effects are different, and so are their magnitudes; in small systems the former is greater by an order of magnutude than the latter. One therefore expects that by analogy with the Hartree-Fock method the exchange part of the functional must be responsible for cancellation of $J^\mathrm{SI}$.

Our goal here is to quantify the extent of this cancellation. In other words, we would like to quantify how well the common KS exchange density functionals model the {\em SI-free} component of the exchange energy. The SI-free part of the exchange has been sometimes referred to in the literature as the {\em interelectronic}\cite{VRG:kaldor:1967:JCP,VRG:lindgren:1972:PRA} or {\em genuine}\cite{VRG:kutzelnigg:2003:Ecwficap} exchange. Although the total ({\em gross}) exchange energy $E_{\mathrm{x}}^{\mathrm{KS}}$ is orbital-invariant, the SI-free part of $E_{\mathrm{x}}^{\mathrm{KS}}$,
\begin{align}
\label{eq:EbarxKS}
\bar{E}_{\mathrm{x}}^{\mathrm{KS}} = & E_{\mathrm{x}}^{\mathrm{KS}} + J^\mathrm{SI},
\end{align}
is not, due to the orbital variance of $E^\mathrm{SI}$. For example, in the canonical basis the self-interaction energy of the homogeneous electron gas is zero (due to the infinite extent of plane-waves), hence $\bar{E}_{\mathrm{x}}^{\mathrm{KS}} = E_{\mathrm{x}}^{\mathrm{KS}}$. However, in a localized (e.g., Wannier\cite{VRG:wannier:1937:PR}) basis the self-interaction energy is nonzero and the genuine exchange energy is less than the total exchange energy.\cite{VRG:norman:1983:PRB}
Thus we must first come up with a unique definition for $\bar{E}_{\mathrm{x}}^{\mathrm{KS}}$. Second, we need to define a method for obtaining reference values of genuine exchange in atoms and molecules.
And since an accurate model of $\bar{E}_{\mathrm{x}}^{\mathrm{KS}}$ may come as a result of cancellation of errors between different orbitals, we will want to examine individual orbital contributions to the genuine exchange energy.

The exchange energy used as the reference (``exact'') target for KS exchange is usually taken to be the exchange contribution to the Hartree-Fock energy, $E_\mathrm{x}^\mathrm{HF}$ (\cref{eq:ExHF}). Since $E_\mathrm{x}^\mathrm{HF}$ includes SI, it is reasonable to take the SI-free part of $E_\mathrm{x}^\mathrm{HF}$ as the target for SI-free KS exchange. However, since the HF orbitals are determined self-consistently in presence of exchange interactions, the more appropriate way to define the {\em genuine} exchange energy is as the difference between the ``exact'' exchange-including correlation-free mean field model (HF) and the exchange-free self-interaction-free mean-field model.
The appropriate choice for the latter is the (orthogonal) Hartree method.
To motivate such choice consider the SI-free parts of the Coulomb and exchange Hartree-Fock energies:
\begin{align}
\label{eq:JH}
  J^{\rm H} \equiv & J - J^\mathrm{SI} = \frac{1}{2} \sum_{i \neq j} \braket{ij}{ij} = \sum_{i < j} \braket{ij}{ij}, \\
\label{eq:EbarxHF}
  \bar{E}_\mathrm{x}^{\rm HF} \equiv & E_\mathrm{x}^\mathrm{HF} + J^\mathrm{SI} = - \frac{1}{2} \sum_{i \neq j} \braket{ji}{ij} = -\sum_{i < j} \braket{ji}{ij}.
\end{align}
The restricted summation form of \cref{eq:JH,eq:EbarxHF} are shown to underscore the lack of SI in these quantities.
Note that {\em canceling} the self-repulsion in the exchange energy means {\em adding} the self-repulsion energy to it. Although $J^\mathrm{SI}$ is positive, $\bar{E}_\mathrm{x}^{\rm HF}$ is negative for any choice of HF orbitals, because $\forall i,j: \braket{ji}{ij} > 0$ due to the positivity of the Coulomb operator.

Lastly, for any fixed choice of orbitals the SI-free HF exchange energy has a straightforward orbital decomposition:
\begin{align}
\label{eq:EbarxHF_per_orb}
\bar{E}_\mathrm{x}^\mathrm{HF} = & \sum_i [\bar{E}_\mathrm{x}^\mathrm{HF}]_i \\
\label{eq:EbarxHFi}
[\bar{E}_\mathrm{x}^\mathrm{HF}]_i = & - \frac{1}{2} \sum_{j \neq i} \braket{ji}{ij}.
\end{align}
A particular choice of HF orbitals will be denoted by replacing ``HF'' with the label denoting the particular orbital type (``c'' for canonical, ``FB'' for Foster-Boys, ``ER'' for Edmiston-Ruedenberg). Just like $\bar{E}_\mathrm{x}^\mathrm{HF}$, $[\bar{E}_\mathrm{x}^\mathrm{HF}]_i$ is negative for any choice of orbitals.

\subsection{Genuine Exchange and Orthogonal Hartree Method}
\label{subsection:genuine-exchange-orthogonal-hartree}

$J^{\rm H}$ and $\bar{E}_\mathrm{x}^{\rm HF}$ can be thought of as the genuine Coulomb and exchange energies in the Hartree-Fock method, thereby giving the HF energy (\cref{eq:EHF}) in the {\em SI-free} form:
\begin{align}
\label{eq:EHF-sifree}
E^\mathrm{HF} = & \sum_i \bra{i}\hat{h}\ket{i} + J^\mathrm{H} + \bar{E}_\mathrm{x}^\mathrm{HF}.
\end{align}
Omission of the genuine exchange energy in \cref{eq:EHF-sifree} leads to the energy of the Hartree method:\cite{VRG:hartree:1928:MPCPS,VRG:hartree:1928:MPCPSa}
\begin{align}
\label{eq:EH}
E^\mathrm{H} = & \sum_i \bra{i}\hat{h}\ket{i} + J^\mathrm{H}.
\end{align}
Note that the two-electron part of the Hartree energy is $J^\mathrm{H}$ and not $J$, thus denoting $J$ as the ``Hartree energy", as done throughout much of the KS DFT literature,\cite{VRG:engel:2011:} is improper.

Note that the Hartree method, unlike Hartree-Fock, treats electrons as \textit{distinguishable}. However, if we insist on the orbitals being orthonormal (which was not done in the original work of Hartree) and limit the (spin) orbitals to at most single occupancy, the orthogonal Hartree method will produce 1-particle density matrices that satisfy the fermionic n-representability.
For brevity we will skip the ``orthogonal'' qualifier: henceforth ``Hartree'' will mean ``orthogonal Hartree'' and label ``H'' will denote quantities produced by the orthogonal Hartree method.

Thus genuine exchange energy will be defined as the difference between
the Hartree-Fock and Hartree energies, both determined {\em self-consistently}:
\begin{align}
\label{eq:Ebarx}
\bar{E}_\mathrm{x} = & E^\mathrm{HF} - E^\mathrm{H}.
\end{align}
Although plugging \cref{eq:EHF-sifree,eq:EH} into \cref{eq:Ebarx} may suggest that $\bar{E}_\mathrm{x}$ is equivalent to $\bar{E}_\mathrm{x}^\mathrm{HF}$, this is not the case, due to the differences between the self-consistent Hartree-Fock and Hartree orbitals. Due to the orbital variance of the Hartree energy, the self-consistent Hartree solver differs substantially from the Roothaan-Hall-type Hartree-Fock orbital solvers used commonly. The details of the Hartree orbital solver can be found in \cref{section:solver-hartree}

The difference between the Hartree-Fock and Hartree orbitals makes it impossible to define an orbital decomposition of $\bar{E}_\mathrm{x}$.
As we will demonstrate numerically in \cref{subsection:results-hartree}, in practice $\bar{E}_\mathrm{x}$ is close to $\bar{E}_\mathrm{x}^\mathrm{HF}$ as long as the latter is evaluated with {\em localized} HF orbitals. Namely, the Edmiston-Ruedenberg (ER) localized orbitals,\cite{VRG:edmiston:1963:RMP} as demonstrated numerically in \cref{subsection:results-hartree} and argued formally in \cref{section:solver-hartree}, approximate the Hartree orbitals very closely.
This idea has also been mentioned by Kutzelnigg
without numerical evidence.\cite{VRG:kutzelnigg:2003:Ecwficap}
Therefore to support orbital decomposition of the genuine exchange energy, instead of using {\em exact} genuine exchange \cref{eq:Ebarx} as the reference we will use HF genuine exchange $\bar{E}_\mathrm{x}^\mathrm{HF}$ evaluated with HF ER orbitals.

\subsection{Exact Conditions on Genuine Exchange}
\label{subsection:exact-conditions-on-genuine-exchange}

Genuine exchange energies satisfy two useful exact conditions:
\begin{enumerate}
    \item For a 1-electron system the genuine exchange energy (evaluated exactly or using ER orbitals) must be zero:
\begin{align}
\label{eq:Ebar-1e}
\bar{E}_\mathrm{x} \overset{1e}{=} 0.
\end{align}
\item For a system with two same-spin electrons in orbitals $0$ and $1$ (e.g., He atom in triplet state, Be atom, etc.) the genuine exchange energy must be the 
same for each orbital:
\begin{align}
\label{eq:Ebar-2e}
[\bar{E}_\mathrm{x}^\mathrm{HF}]_0 \overset{2e}{=} - \frac{1}{2} \braket{10}{01} =  [\bar{E}_\mathrm{x}^\mathrm{HF}]_1.
\end{align}
\end{enumerate}

\subsection{Orbital Decomposition of KS Exchange}
\label{subsection:orbital-decomposition-ks-exchange}

To be able to compare the per-orbital genuine exchange energies between the reference and its values obtained with DFT approximantions it is yet again necessary to remove the effects of self-consistency on the orbital space. Hence we will use HF ER orbitals for evaluation of the genuine exchange at both the HF and KS levels. The decomposition of the KS gross exchange energy follows \cref{eq:Exc_by_orb}:
\begin{align}
  \label{eq:Ex_by_orb}
E_{\mathrm{x}}^{\mathrm{KS}} = & \sum_i \int v_\mathrm{x}^\mathrm{KS}[n](\mathbf{r}) \, n_i(\mathbf{r}) \, \mathrm{d}\mathbf{r} \equiv \sum_i [E_{\mathrm{x}}^{\mathrm{KS}}]_i, \\
[E_{\mathrm{x}}^{\mathrm{KS}}]_i = & \int v_\mathrm{x}^\mathrm{KS}[n](\mathbf{r}) \, n_i(\mathbf{r}) \, \mathrm{d}\mathbf{r}.
\end{align}
The per-orbital genuine exchange is obtained by canceling the orbital self-repulsion:
\begin{align}
  \label{eq:EbarxKSi}
[\bar{E}_{\mathrm{x}}^{\mathrm{KS}}]_i = & [E_{\mathrm{x}}^{\mathrm{KS}}]_i + \frac{1}{2} \braket{ii}{ii} = \int \bar{v}_\mathrm{x}^\mathrm{KS}[n](\mathbf{r}) \, n_i(\mathbf{r}) \, \mathrm{d}\mathbf{r},
\end{align}
where $\bar{v}_\mathrm{x}^\mathrm{KS}$ is the genuine KS exchange potential,
\begin{align}
\bar{v}_\mathrm{x}^\mathrm{KS}[n](\mathbf{r}) = v_\mathrm{x}^\mathrm{KS}[n](\mathbf{r}) + \frac{1}{2} v_{J}[n_i](\mathbf{r}),
\end{align}
and $v_{J}$ is the Coulomb potential of a density:
\begin{align}
  v_J[n](\mathbf{r}) = \int \frac{n(\mathbf{r}')}{|\mathbf{r} - \mathbf{r}'|} \, \mathrm{d}\mathbf{r}'.
\end{align}

Standard KS DFT functionals satisfy neither of the genuine exchange exactness conditions from \cref{subsection:exact-conditions-on-genuine-exchange}. The per-orbital analysis of their failings will be discussed in \cref{section:results-orb-anatomy}.

\subsection{Connection to Perdew-Zunger Self-Interaction Correction}
\label{subsection:connection-pz-sic}

For a 1-electron system the genuine KS exchange potential should be zero. Since in such system $n=n_0$, the exact functional should produce an exchange potential that satisfies
\begin{align}
\label{eq:vx-exact-1e}
    v_\mathrm{x}^\mathrm{exact\ KS}[n] \overset{1e}{=} - \frac{1}{2} v_{J}[n].
\end{align}
And since there is no correlation in a 1-electron system, the same argument applies to the xc potential as well:
\begin{align}
\label{eq:vxc-exact-1e}
    v_\mathrm{xc}^\mathrm{exact\ KS}[n] \overset{1e}{=} - \frac{1}{2} v_{J}[n].
\end{align}
Perdew and Zunger proposed\cite{VRG:perdew:1981:PRB} to {\em correct} the $v_\mathrm{xc}^\mathrm{KS}[n]$ potential felt by each orbital in \cref{eq:Exc_by_orb} as follows:
\begin{align}
    v_\mathrm{xc}^\mathrm{PZ}[n,n_i] \equiv v_\mathrm{xc}^\mathrm{KS}[n] - v_\mathrm{xc}^\mathrm{KS}[n_i] - \frac{1}{2} v_\mathrm{J}[n_i]
\end{align}
This ensures that for any 1-electron system ($n=n_0$) the PZ xc potential satisfies \cref{eq:vxc-exact-1e} and is thus exact
\begin{align}
    v_\mathrm{xc}^\mathrm{PZ}[n,n] \overset{1e}{=} - \frac{1}{2} v_{J}[n].
\end{align}
As a result PZ-corrected functionals satisfy the 1-electron genuine exchange exactness condition, \cref{eq:Ebar-1e}. However, it is easy to see that they do not satisfy the 2-electron condition, \cref{eq:Ebar-2e}. For many-electron systems the PZ-corrected KS DFT functionals are also not exact. While for the local density approximation the use of PZ correction produces substantial improvement, that's not always the case for more accurate KS functionals.\cite{VRG:perdew:2015:AIAMaOP} The orbital anatomy of exchange energies will allow us to quantify the performance of the PZ-corrected functionals in more detail.

\section{Technical Details}
\label{section:tech-det-orb-anatomy}

Calculations were performed with a modified version of the Hartree-Fock program available in Libint2.\cite{VRG:valeev:2020:}
The GNU Scientific Library\cite{VRG:galassi:2009:} was used for determining step sizes in the line search used in the solution of the Hartree orbitals.
All integrals were evaluated using Libint2, and density functionals were evaluated using LibXC (version 5.0.0),\cite{VRG:lehtola:2018:S} through the ExchCXX package.\cite{VRG:williams:2021:achieving}
Quadrature was done with Becke's method\cite{VRG:becke:1988:JCPa} and using Gau2Grid and IntegratorXX, with integration mesh of 300 radial Euler Maclaurin \cite{VRG:murray:1993:MP} points, 1202 angular Lebedev-Laikov \cite{VRG:lebedev:1999:DM} for all atoms. All calculations used cc-pVTZ\cite{VRG:dunning:1989:JCP,VRG:prascher:2011:TCA,VRG:woon:1993:JCP,VRG:woon:1994:JCP} as the orbital basis with density fitting applied with cc-pVTZ-RI.\cite{VRG:weigend:2002:JCP,VRG:hattig:2005:PCCP}
Density functionals used were Slater exchange (LDAx),\cite{VRG:dirac:1930:MPCPS} B88,\cite{VRG:becke:1988:PRA} exchange-only PBE (PBEx),\cite{VRG:perdew:1996:PRL} exchange-only revised PBE (revPBEx),\cite{VRG:zhang:1998:PRL} exchange-only (TPSSx),\cite{VRG:tao:2003:PRL} exchange-only revised TPSS (revTPSSx),\cite{VRG:perdew:2009:PRL,VRG:perdew:2011:PRL} exchange-only SCAN (SCANx),\cite{VRG:sun:2015:PRL} exchange-only revised SCAN (revSCANx),\cite{VRG:mezei:2018:JCTC} and exchange-only M06-L (M06-Lx).\cite{VRG:zhao:2006:JCP}

The Massively Parallel Quantum Chemistry (MPQC) version 4 program package,\cite{VRG:peng:2020:JCP} was used as a C++ library to perform restricted Hatree-Fock calculations to prepare orbitals for evaluation of KS and beyond KS functionals.
All atomic calculations were performed with restricted open-shell direct minimization formalism (not using MPQC).
The Jacobi solver\cite{VRG:boys:1960:RMP} of the ER localizer was stopped when the max rotation angle was less than $10^{-7}$ and $10^{-10}$ radians for atoms and molecules, respectively (note that some singly-occupied orbitals in atoms, such as p-orbitals in N and P atoms, cannot be localized).
Structures for the small molecules used came from the Gaussian output files provided by NIST online, \cite{VRG:johnson:2002:} using the optimization method and basis set for the G2 set.\cite{VRG:curtiss:1997:JCP}

\section{Results}
\label{section:results-orb-anatomy}

\subsection{Genuine Exchange: Exact vs Hartree-Fock-based Definitions}
\label{subsection:results-hartree}

\begin{figure}
    \centering
    \includegraphics[width=0.9\columnwidth]{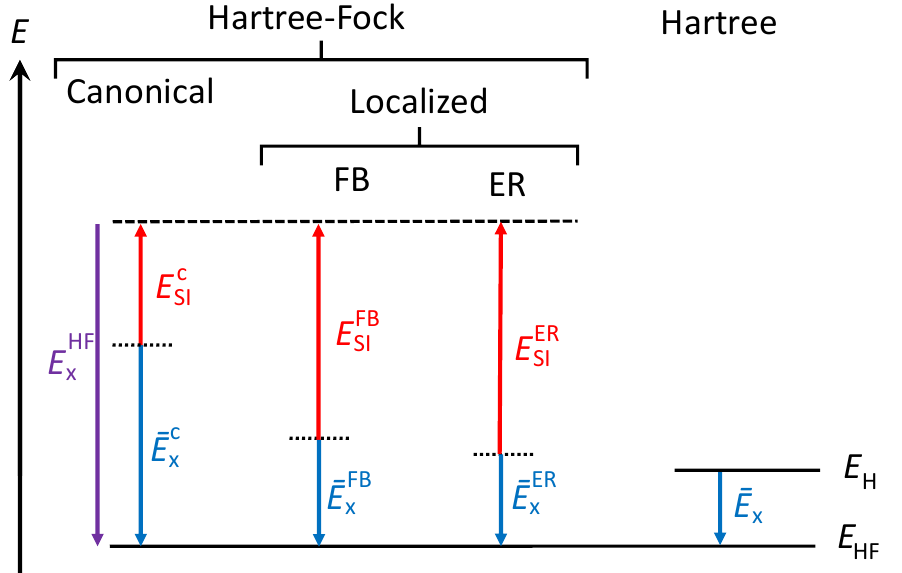}
    \caption{Schematic representation (not to scale) of HF exchange energy broken into the self-interaction and genuine exchange components (see \cref{eq:EbarxHF}), and its relationship to the {\em exact} genuine exchange (see \cref{eq:Ebarx}).}
    \label{fig:HF_v_H_scheme}
\end{figure}

\Cref{table:HF_vs_H,table:genex_errors} report the genuine exchange energies evaluated exactly as the difference between Hartree-Fock and Hartree energies (\cref{eq:Ebarx}) and extracted from the Hartree-Fock orbitals (\cref{eq:EbarxHF}) for three choices of orbitals: canonical, Foster-Boys,\cite{VRG:foster:1960:RMP,VRG:boys:1960:RMP} and Edmiston-Ruedenberg\cite{VRG:edmiston:1963:RMP}. A detailed breakdown is reported for the representative set of three molecules (containing single, double, and triple covalent bonds) in \cref{table:HF_vs_H}, with a more focused analysis reported for 11 molecules in \cref{table:genex_errors}. Several conclusions can be drawn:
\begin{itemize}
\item Whereas the gross Hartree-Fock exchange energy is $\geq 10\%$ of the total energy, the exact genuine exchange energy is $\leq 1 \%$ of the total. The latter should be easier to model in SI-free KS framework than the former in the standard KS framework.
\item The genuine exchange energy can be accurately approximated using localized Hartree-Fock orbitals via \cref{eq:EbarxHF}. Orbital localization increases the SI energy and thereby reduces the genuine exchange energy. Since Hartree orbitals are guaranteed to produce the lowest energy, the exact genuine exchange energy $\bar{E}_\mathrm{x}$ is therefore guaranteed to be smaller in absolute magnitude than $\bar{E}_\mathrm{x}^\mathrm{HF}$. \Cref{fig:HF_v_H_scheme} contains a schematic depiction of these relationships.
\item Genuine exchange energies obtained from FB orbitals are less accurate than the ER orbitals by roughly 5\%, although for one outlier (Li$_2$) FB-based genuine exchange is almost 2.5 times too large. This calls for caution when using Foster-Boys orbitals for correcting KS DFT for self-interaction.\cite{VRG:li:2018:NSR} The largest deviation of ER-based genuine exchange energy was found to be 14.2\% for CH$_4$.
\item The ER HF genuine exchange energy is on average within 10\% of the exact (Hartree-based) genuine exchange. The excellent agreement reflects the deep connection between orthogonal Hartree and ER as discussed in \cref{section:solver-hartree}. The residual difference between exact and ER HF-derived genuine exchange energies, on the order of only 1\% of the total energy, is due to the difference between Hartree and Hartree-Fock orbital spans.
\end{itemize}

\begin{table}
    \centering
    \begin{tabular}{ldllddd}
    \toprule
    \toprule
    \multicolumn{3}{l}{C$_2$H$_4$} & \multicolumn{4}{c}{Exchange Energy}  \\ \cmidrule(lr){4-7}
    \multicolumn{3}{c}{Total Energy} & {} & \multicolumn{1}{c}{Canonical} & \multicolumn{1}{c}{FB} & \multicolumn{1}{c}{ER} \\ \cmidrule(lr){1-3} \cmidrule(lr){4-7} 
    $E^\mathrm{HF}$ & -78.064 & {} & $E^{\mathrm{HF}}_{\mathrm{x}}$ & -11.745 & -11.745 & -11.745 \\
    $E^\mathrm{H}$ & -77.575 & {} & $E^{\mathrm{SI}}$ & 6.537 & 11.169 & 11.202 \\
    $\bar{E}_{\mathrm{x}}$ & -0.489 & {} & $\bar{E}_{\mathrm{x}}^\mathrm{HF}$ & -5.208 & -0.576 & -0.543 \\
    \multicolumn{6}{c}{} \\
    \hline
    \multicolumn{3}{l}{CO} & \multicolumn{4}{c}{Exchange Energy}  \\ \cmidrule(lr){4-7}
    \multicolumn{3}{c}{Total Energy} & {} & \multicolumn{1}{c}{Canonical} & \multicolumn{1}{c}{FB} & \multicolumn{1}{c}{ER} \\ \cmidrule(lr){1-3} \cmidrule(lr){4-7} 
    $E^\mathrm{HF}$ & -112.777  & {} & $E^{\mathrm{HF}}_{\mathrm{x}}$ & -13.305 & -13.305 & -13.305 \\
    $E^\mathrm{H}$ & -111.911 & {} & $E^{\mathrm{SI}}$ & 11.663 & 12.318 & 12.365 \\
    $\bar{E}_{\mathrm{x}}$ & -0.866 & {} & $\bar{E}_{\mathrm{x}}^\mathrm{HF}$ & -1.642 & -0.987 & -0.940 \\
    \multicolumn{6}{c}{} \\
    \hline
    \multicolumn{3}{l}{HF} & \multicolumn{4}{c}{Exchange Energy}  \\ \cmidrule(lr){4-7}
    \multicolumn{3}{c}{Total Energy} & {} & \multicolumn{1}{c}{Canonical} & \multicolumn{1}{c}{FB} & \multicolumn{1}{c}{ER} \\ \cmidrule(lr){1-3} \cmidrule(lr){4-7} 
    $E^\mathrm{HF}$ & -100.057 & {} & $E^{\mathrm{HF}}_{\mathrm{x}}$ & -10.430 & -10.430 & -10.430 \\
    $E^\mathrm{H}$ & -99.182 & {} & $E^{\mathrm{SI}}$: & 8.803 & 9.451 & 9.475 \\
    $\bar{E}_{\mathrm{x}}$ & -0.875 & {} & $\bar{E}_{\mathrm{x}}^\mathrm{HF}$ & -1.627 & -0.978 & -0.954 \\
    
    \bottomrule
    \bottomrule
    \end{tabular}
    \caption{Approximation of genuine exchange energy($\bar{E}_{\mathrm{x}}$, \cref{eq:Ebarx}) by its Hartree-Fock-derived counterpart ($\bar{E}_{\mathrm{x}}^\mathrm{HF}$, \cref{eq:EbarxHF}) obtained with canonical and localized orbitals for three representative molecules containing single, double, and triple covalent bonds. All values in $E_{\mathrm{h}}$.}
    \label{table:HF_vs_H}
\end{table}

\begin{table}
    \centering
    \begin{tabular}{lddddd}
    \toprule
    \toprule
    System & \multicolumn{1}{c}{$\bar{E}_{\mathrm{x}}$}  & \multicolumn{2}{c}{$\bar{E}_{\mathrm{x}}^\mathrm{HF}$} & \multicolumn{2}{c}{\% Error vs. $\bar{E}_{\mathrm{x}}$} \\ \cmidrule(lr){3-4} \cmidrule(lr){5-6} 
    {} & {} & \multicolumn{1}{c}{FB} & \multicolumn{1}{c}{ER} & \multicolumn{1}{c}{FB} & \multicolumn{1}{c}{ER} \\
    \hline
    HF & -0.875 & -0.978 & -0.954 & -11.8 & -9.1 \\
    H$_2$O & -0.621 & -0.711 & -0.690 & -14.5 & -11.1 \\
    H$_2$O$_2$ & -1.150 & -1.300 & -1.254 & -13.1 & -9.1 \\
    C$_2$H$_4$ & -0.489 & -0.576 & -0.543 & -17.8 & -11.2 \\
    C$_2$H$_2$ & -0.527 & -0.614 & -0.579 & -16.5 & -10.0 \\
    NH$_3$ & -0.411 & -0.482 & -0.464 & -17.5 & -13.0 \\
    Li$_2$ & -0.007 & -0.023 & -0.007 & -243.5 & -5.7 \\
    CH$_4$ & -0.236 & -0.285 & -0.270 & -20.5 & -14.2 \\
    N$_2$ & -0.812 & -0.922 & -0.877 & -13.6 & -8.1 \\
    CO & -0.866 & -0.987 & -0.940 & -14.0 & -8.6 \\
    H$_2$CO & -0.834 & -0.956 & -0.913 & -14.7 & -9.5 \\

    \bottomrule
    \bottomrule
    \end{tabular}
    \caption{Approximation of genuine exchange energy ($\bar{E}_{\mathrm{x}}$) using localized HF orbitals for a set of 11 molecules from the G2 set. Energies in $E_{\mathrm{h}}$.}
    \label{table:genex_errors}
\end{table}
Clearly canonical HF orbitals are not suitable to approximate $\bar{E}_{\mathrm{x}}$.
Although the Foster-Boys (FB) localization scheme\cite{VRG:foster:1960:RMP,VRG:boys:1960:RMP} performs much better, it is generally 5\% worse than ER, as shown in \cref{table:genex_errors}.

\begin{figure}
    \centering
    \includegraphics[width=0.9\columnwidth]{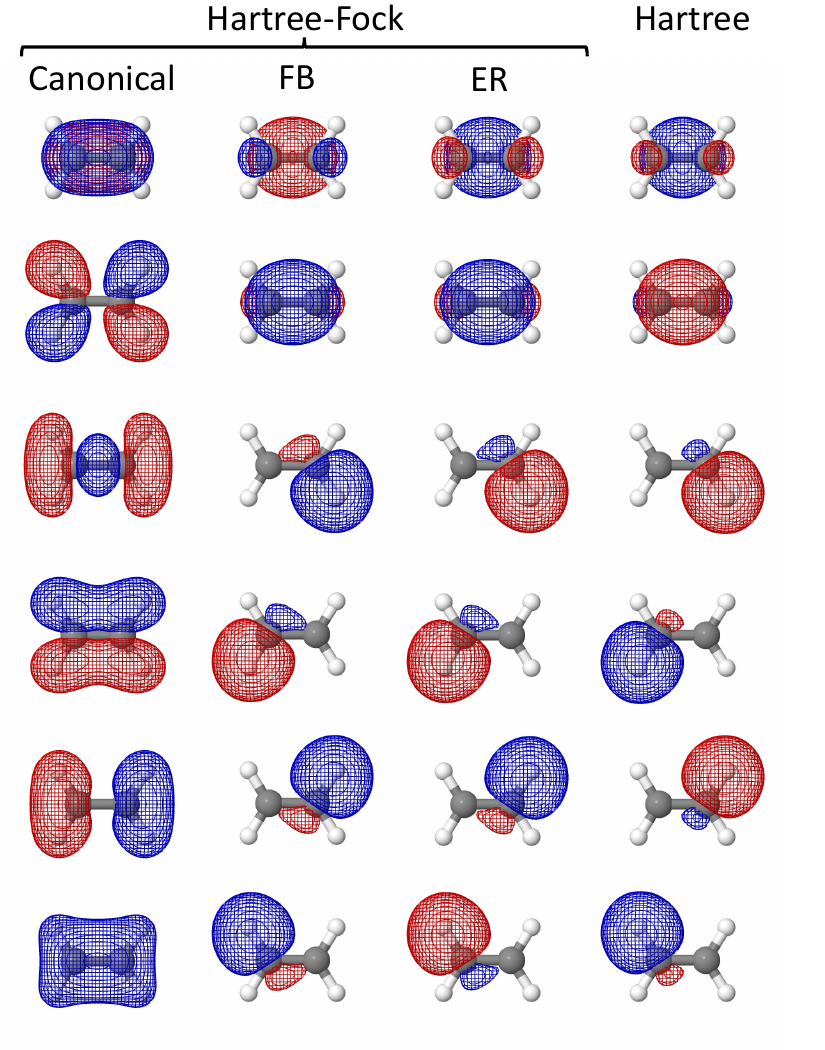}
    \caption{Valence orbitals for C$_2$H$_4$ optimized with Hartree-Fock (both canonical and two versions of localized) and Hartree methods.}
    \label{fig:hp_v_hf}
\end{figure}
\Cref{fig:hp_v_hf} visually illustrates the differences between Hartree-Fock and Hartree orbitals for the ethylene molecule.
Both choices of localized HF orbitals are similar to the Hartree orbitals (modulo the arbitrary phase).
There are two types of localized orbitals here: $\sigma$-bonding and ``banana'' bonding.
The normal $\pi$-bonding and $\sigma$-bonding orbitals expected in an alkene are mixed to form the ``banana'' bonds.\cite{VRG:kaldor:1967:JCP}
The ``banana'' bonds are bending either into or out of the plan of the page in \cref{fig:hp_v_hf}.
The only minor difference between FB, ER, and Hartree orbitals can be found in the ``minor'' lobe of the $\sigma$-bonding orbitals: the lobe decreases in this sequence.

\subsection{Orbital Breakdown of KS Exchange Energies}
\label{subsection:results_ks-dft}

As mentioned, to enable orbital decomposition of energies the effects of self-consistency are neglected and HF ER orbitals (as closely approximating the Hartree orbitals) were used to analyze the gross and genuine exchange energy contributions for Hartree-Fock and a variety of KS DFT functionals.

Total genuine exchange energies for HF and KS models are defined in \cref{eq:EbarxHF} and \cref{eq:EbarxKS}, with the corresponding orbital components given in \cref{eq:EbarxHFi} and \cref{eq:EbarxKSi}, respectively. The errors in per-orbital KS genuine exchange energies are defined with respect to the HF reference,
\begin{align}
\label{eq:DEbarxKS}
  \Delta [\bar{E}_\mathrm{x}^\mathrm{KS}]_i \equiv [\bar{E}_\mathrm{x}^\mathrm{KS}]_i - [\bar{E}_\mathrm{x}^\mathrm{HF}]_i.
\end{align}
Since the exact exchange component of any global hybrid functional
does not contribute to the genuine exchange error (total or per-orbital), then the genuine exchange errors of global hybrids are identical to the genuine exchange error of the semilocal component of the functional. In other words, the genuine exchange error of a global hybrid including $x\%$ of the exact exchange is obtained by scaling the error of its semilocal component by $1-x/100$.

\subsubsection{Atoms}
\label{subsubsection:results_ks-dft_atoms}

Orbital decomposition of HF and KS exchange energies for Ne and Ar atoms is presented in 
\cref{table:atom_per-orbital}, with data for the rest of the atoms with Z $\leq$ 18 located in the Supplementary Information.

\begin{table*}
    \centering
    \begin{tabular}{lccccccccc}
    \toprule
    \toprule
    Ne & {} & \multicolumn{4}{c}{Gross Exchange} & \multicolumn{4}{c}{Genuine Exchange} \\ \cmidrule(lr){3-6} \cmidrule(lr){7-10} 
    orbital & $E_{\mathrm{SI}}$ & HF & LDAx & B88 & PBEx & HF & LDAx & B88 & PBEx \\
    \hline

    1s & 6.137 & -6.267 & -5.475 & -6.075 & -6.043 & -0.130 & 0.663 & 0.062 & 0.094 \\ 
    2sp$^3$ & 1.180 & -1.462 & -1.390 & -1.516 & -1.507 & -0.282 & -0.211 & -0.337 & -0.327 \\ 
    2sp$^3$ & 1.180 & -1.462 & -1.390 & -1.516 & -1.507 & -0.282 & -0.211 & -0.337 & -0.327 \\ 
    2sp$^3$ & 1.180 & -1.462 & -1.390 & -1.516 & -1.507 & -0.282 & -0.211 & -0.337 & -0.327 \\ 
    2sp$^3$ & 1.180 & -1.462 & -1.390 & -1.516 & -1.507 & -0.282 & -0.211 & -0.337 & -0.327 \\ 
    \hline
    totals: & 10.856 & -12.113 & -11.036 & -12.140 & -12.069 & -1.257 & -0.180 & -1.284 & -1.213 \\ 
    \multicolumn{10}{c}{} \\
    Ar & {} & \multicolumn{4}{c}{Gross Exchange} & \multicolumn{4}{c}{Genuine Exchange} \\ \cmidrule(lr){3-6} \cmidrule(lr){7-10} 
    orbital & $E_{\mathrm{SI}}$ & HF & LDAx & B88 & PBEx & HF & LDAx & B88 & PBEx \\
    \hline
    1s & 11.384 & -11.812 & -10.405 & -11.362 & -11.301 & -0.428 & 0.979 & 0.022 & 0.083 \\ 
    2sp$^3$ & 2.937 & -3.706 & -3.525 & -3.781 & -3.763 & -0.769 & -0.589 & -0.844 & -0.827 \\ 
    2sp$^3$ & 2.937 & -3.706 & -3.525 & -3.781 & -3.763 & -0.769 & -0.589 & -0.844 & -0.827 \\ 
    2sp$^3$ & 2.937 & -3.706 & -3.526 & -3.781 & -3.763 & -0.769 & -0.589 & -0.844 & -0.827 \\ 
    2sp$^3$ & 2.937 & -3.706 & -3.525 & -3.781 & -3.763 & -0.769 & -0.589 & -0.844 & -0.826 \\ 
    3sp$^3$ & 0.696 & -0.882 & -0.839 & -0.917 & -0.911 & -0.186 & -0.143 & -0.221 & -0.215 \\ 
    3sp$^3$ & 0.696 & -0.882 & -0.839 & -0.917 & -0.911 & -0.186 & -0.143 & -0.221 & -0.215 \\ 
    3sp$^3$ & 0.696 & -0.882 & -0.839 & -0.917 & -0.911 & -0.186 & -0.143 & -0.221 & -0.215 \\ 
    3sp$^3$ & 0.696 & -0.882 & -0.839 & -0.917 & -0.911 & -0.186 & -0.143 & -0.221 & -0.215 \\
    \hline
    totals: & 25.915 & -30.164 & -27.863 & -30.154 & -29.996 & -4.249 & -1.948 & -4.238 & -4.081 \\ 

    \bottomrule
    \bottomrule
    \end{tabular}
    \caption{Breakdown of exchange energy by orbital for Ne and Ar atoms. All energies in $E_{\mathrm{h}}$.}
    \label{table:atom_per-orbital}
\end{table*}
The first thing to notice is that because the orbitals we use are ER localized HF orbitals, we do not have the usual atomic structure (1s, 2s, 2p, etc.).
Instead, we have the core (basically 1s) and the valence is hybridized as equivalent sp$^3$ orbitals.

In agreement with the findings in \cref{subsection:results-hartree}, the bulk of gross exchange energy (HF or KS) for every orbital accounts for canceling of its contribution to the self-interaction energy. The latter is inversely related to the extent of the orbital, hence for core orbitals the self-interaction and the gross exchange are much greater in magnitude than those of the valence orbitals.
However, contributions to the genuine exchange energy from the core and valence orbitals are of comparable magnitude. In Ne the core orbital contributes only half as much to the genuine exchange compared to a valence orbital; since there are more valence orbitals, the total is dominated by the latter. In the case of Ar the largest contributions to the genuine exchange come from the outer core (inner valence) orbitals with $n=2$, and with the valence and inner core orbital contribution both substantially smaller.

Let's now consider the performance KS DFT functionals for the exchange energy evaluation, starting with the simplest, LDAx. As is well known, it somewhat underestimates the gross exchange energy. However, its failings become much more pronounced when viewed in the context of genuine exchange: LDAx predicts total genuine exchange energies underestimated by a factor of 7 for Ne and 2.2 for Ar.
Orbital decomposition of energies is highly insightful here.
It reveals, for example, that the LDAx exchange energy contribution of the 1s core orbitals in both Ne and Ar is so severely underestimated that it is not sufficient to cancel the entire self-repulsion!
This leads to a \textit{positive} contribution from the 1s orbital to the genuine exchange energy (0.663 $E_{\mathrm{h}}$ in Ne, 0.979 $E_{\mathrm{h}}$ in Ar); this is physically unreasonable, since the genuine exchange energies (total or orbital-specific) must be negative.
The performance of LDAx for valence orbitals is much better, capturing roughly 75\% of the genuine exchange energy (but it still underestimates the genuine exchange effects).
Overall, we see that the known underestimation of the gross exchange energy by LDAx is largely due to the core orbital.

The gross and genuine exchange energies predicted by the GGA functionals are radically more accurate than LDAx. The orbital breakdown, however, reveals how the accuracy of the total GGA exchange energies is due to the significant cancellation of errors between core and valence orbitals!
Like LDAx, both GGAs underestimate the core orbital exchange energy so much that the genuine exchange contributions from the 1s orbitals are again positive!
These errors are luckily compensated by the overly large exchange energy contributions of the valence orbitals.
To better demonstrate this error cancellation \cref{table:atom_per-orbital-error_ks} reports errors of genuine KS exchange contributions relative to their HF counterparts.
It is important to note that the B88 functional was parameterized to accurately reproduce the exchange energies of some atoms.\cite{VRG:becke:1988:PRA}
Thus, it is not really surprising that the total values are in good agreement with HF, but it is interesting to see how the orbital structure of the contributions does not match HF.
Also, PBEx (which is not fitted to data, and is considered ``non-empirical'') also demonstrates a similar error cancellation between the core and valence orbitals.

To illustrate the extent to which the cancellation of exchange energy errors between different orbitals contributes to the overall accuracy of the KS exchange energy, consider the following plots of exchange energies for atoms $Z\leq 18$:
\begin{itemize}
\item \cref{fig:atoms_total_ex_error} displays the errors in the total KS genuine exchange energies. Not surprisingly, the LDAx exchange energies are significantly less accurate than the GGA counterparts.
\begin{figure}
    \centering
    \includegraphics[width=0.9\columnwidth]{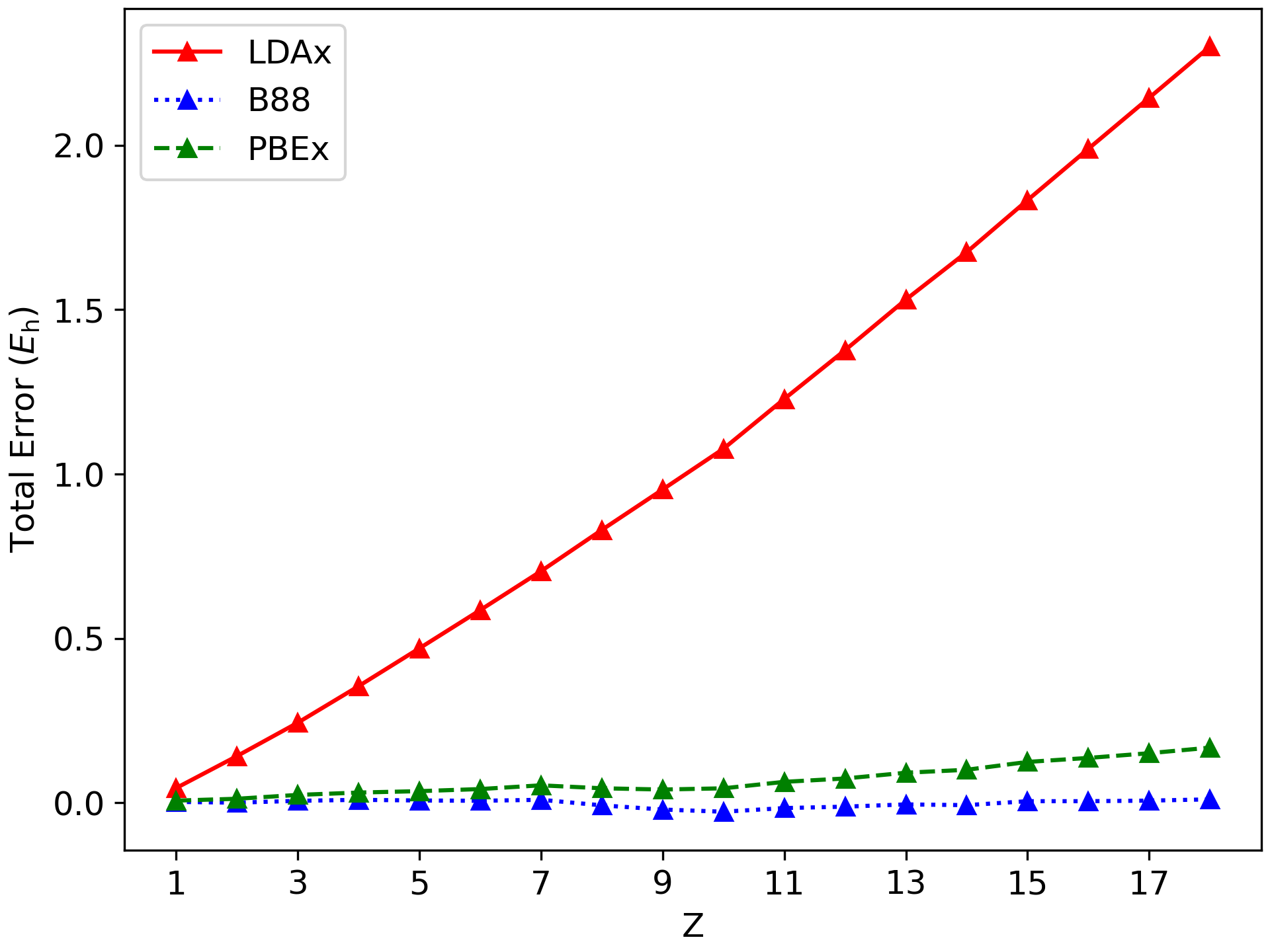}
    \caption{Errors of KS genuine exchange energies for atoms $Z \leq 18$.
    } 
    \label{fig:atoms_total_ex_error}
\end{figure}
\item \cref{fig:atoms_sum_abs_ex_error} displays the sum of absolute errors in per-orbital KS genuine exchange energies. Surprisingly, the LDAx per-orbital exchange errors are only modestly less accurate than the GGA counterparts.
\begin{figure}
    \centering
    \includegraphics[width=0.9\columnwidth]{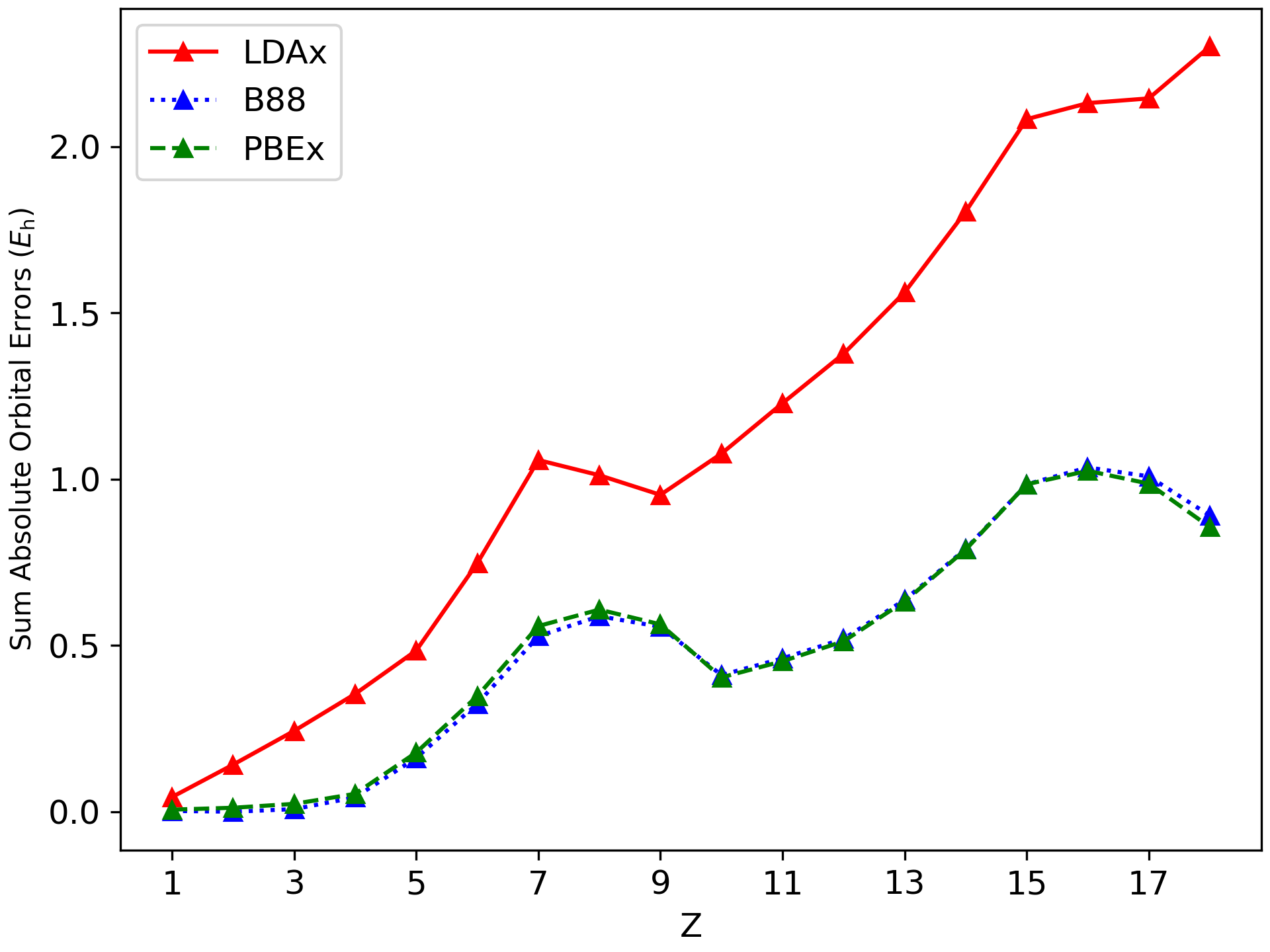}
    \caption{Sum of absolute errors of KS per-orbital genuine exchange energies for atoms $Z \leq 18$.} 
    \label{fig:atoms_sum_abs_ex_error}
\end{figure}
\item To quantify the ``degree'' of error cancellation \cref{fig:atoms_error_cancel} displays the error cancellation measure, defined as
\begin{align}
\label{eq:err-cancellatior-ratio}
  C \equiv 1 - \frac{\vert \sum_i \Delta [\bar{E}_\mathrm{x}^\mathrm{KS}]_i \vert}{\sum_i \vert \Delta [\bar{E}_\mathrm{x}^\mathrm{KS}]_i \vert}.
\end{align}
$C=0$ indicates that no error cancellation is occurring between different orbitals (all per-orbital errors are of the same sign), while $C=1$ means that the per-orbital errors cancel out perfectly (i.e., the total error is small but individual orbital errors are large).
Clearly, LDAx almost never benefits from the error cancellation since it almost always underestimates the magnitude of the exchange energy.
Both GGA functionals clearly benefit from the error cancellation, with the extent of cancellation particularly striking for B88.
\begin{figure}
    \centering
    \includegraphics[width=0.9\columnwidth]{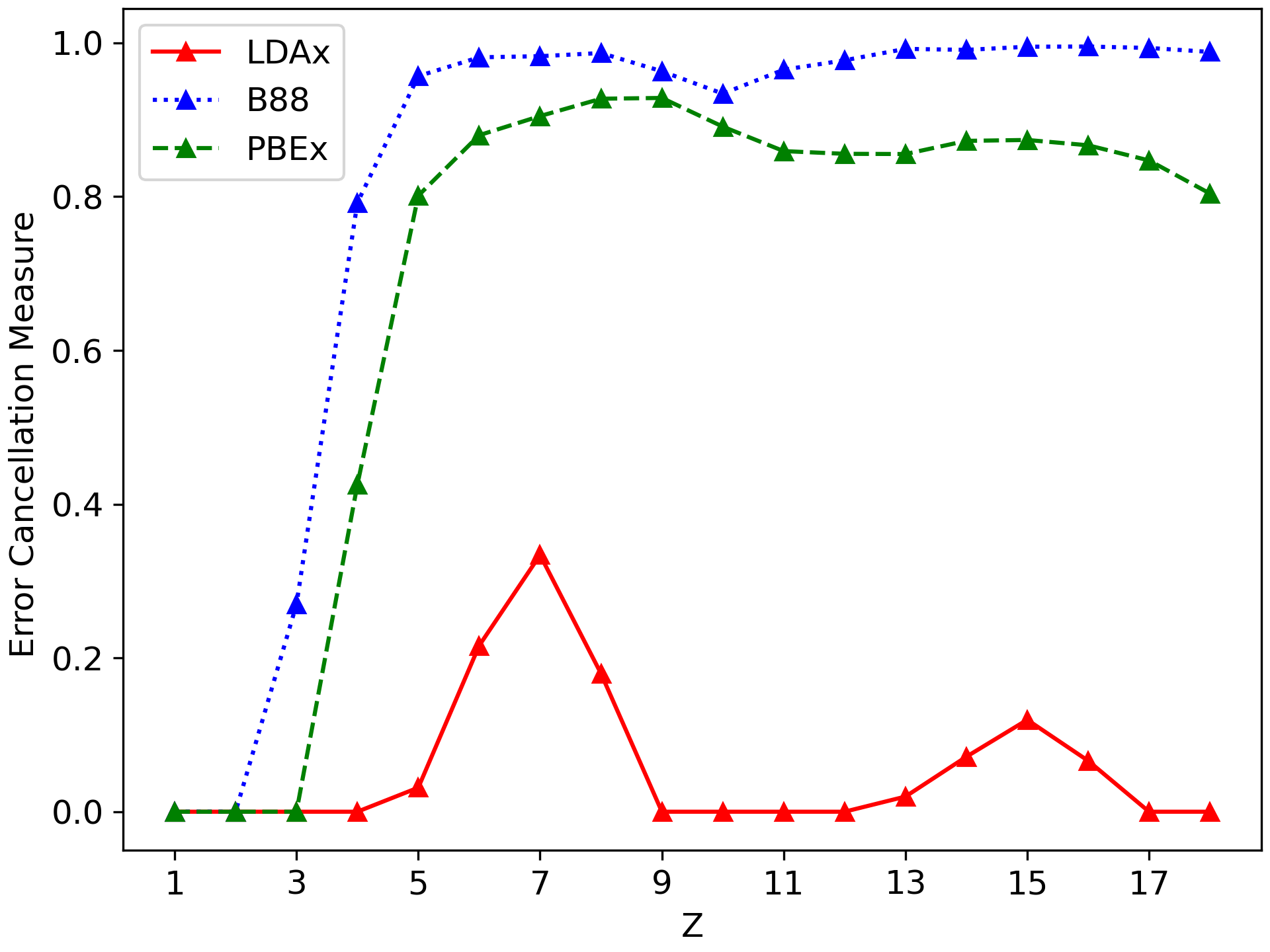}
    \caption{The error cancellation measure (\cref{eq:err-cancellatior-ratio}) for KS genuine exchange energies of atoms $Z \leq 18$.}
    \label{fig:atoms_error_cancel}
\end{figure}
\end{itemize}

\cref{table:atom_per-orbital-error_ks} presents the errors in per-orbital genuine exchange energies for the same two atoms (Ne and Ar)
using three meta-GGAs (revTPSSx, revSCANx, and M06-Lx). 
Note that for KS it does not matter if we compare gross KS to gross HF or genuine KS to genuine HF since both differ by a simple shift (this is not the case for PZ).

The mGGAs total exchange energies are not necessarily systematically better than the GGAs. In the case of Ne, revTPSSx does very well compared to either GGA, while revSCANx is worse.
But for Ar all mGGAs are worse than B88 but better than PBEx.
However, the orbital analysis is again quite revealing, as the accuracy of mGGAs orbital exchange energies are significantly more accurate than the GGA counterparts. The higher accuracy for each orbital unfortunately does not necessarily translate into more accurate total energies due to a less-systematic (but still present) cancellation of errors than in the GGAs.

The use of global hybrids will not affect the error cancellation, but will reduce the individual per-orbital genuine exchange errors. Since the global hybrids designed for general-purposes uses have a modest ($<30\%$) percentage of exact exchange, the per-orbital genuine exchange errors of hybrid GGAs will still be larger than those of (nonhybrid) meta-GGAs.

\begin{table*}
    \centering
    \begin{tabular}{lrrrrrr}
    \toprule
    \toprule
    Ne & \multicolumn{6}{c}{} \\
    orbital & LDAx & B88 & PBEx & revTPSSx & revSCANx & M06-Lx \\
    \hline
    1s & 0.793 & 0.192 & 0.224 & 0.087 & 0.054 & 0.109 \\ 
    2sp$^3$ & 0.071 & -0.055 & -0.045 & -0.023 & -0.027 & -0.022 \\ 
    2sp$^3$ & 0.071 & -0.055 & -0.045 & -0.023 & -0.027 & -0.022 \\ 
    2sp$^3$ & 0.071 & -0.055 & -0.045 & -0.023 & -0.027 & -0.022 \\ 
    2sp$^3$ & 0.071 & -0.055 & -0.045 & -0.023 & -0.027 & -0.022 \\ 
    \hline
    total & 1.077 & -0.027 & 0.044 & -0.004 & -0.055 & 0.020 \\ 
    \multicolumn{7}{c}{} \\
    Ar & \multicolumn{6}{c}{} \\
    orbital & LDAx & B88 & PBEx & revTPSSx & revSCANx & M06-Lx \\
    \hline
    1s & 1.408 & 0.451 & 0.512 & 0.248 & 0.121 & 0.247 \\ 
    2sp$^3$ & 0.181 & -0.075 & -0.057 & -0.009 & -0.045 & -0.013 \\ 
    2sp$^3$ & 0.181 & -0.075 & -0.057 & -0.009 & -0.046 & -0.013 \\ 
    2sp$^3$ & 0.181 & -0.075 & -0.057 & -0.010 & -0.046 & -0.013 \\ 
    2sp$^3$ & 0.181 & -0.075 & -0.057 & -0.009 & -0.045 & -0.013 \\
    3sp$^3$ & 0.043 & -0.035 & -0.029 & -0.014 & -0.009 & -0.009 \\ 
    3sp$^3$ & 0.043 & -0.035 & -0.029 & -0.014 & -0.009 & -0.009 \\ 
    3sp$^3$ & 0.043 & -0.035 & -0.029 & -0.014 & -0.009 & -0.009 \\ 
    3sp$^3$ & 0.043 & -0.035 & -0.029 & -0.014 & -0.009 & -0.009 \\ 
    \hline
    total & 2.301 & 0.010 & 0.168 & 0.152 & -0.097 & 0.158 \\ 
    \bottomrule
    \bottomrule
    \end{tabular}
    \caption{Breakdown of errors in KS genuine exchange energy by orbital for atoms Ne and Ar. All energies in $E_{\mathrm{h}}$.}
    \label{table:atom_per-orbital-error_ks}
\end{table*}

\subsubsection{Molecules}
\label{subsubsection:results_ks-dft_molecules}

A detailed per-orbital breakdown of the gross and genuine HF, LDA, and GGA exchange energies for the three representative molecules considered in \cref{subsection:results-hartree} is presented in \cref{table:molecule_per-orbital} (the rest of the data can be found in the Supplementary Information). Although there is a greater variety of types of orbitals in molecules than in atoms, the trends largely follow those found for the atoms (\cref{table:atom_per-orbital}). LDA underestimates the exchange energies and predicts unphysical (positive) genuine exchange energies, not only for core orbitals but also sometimes for valence orbitals (e.g., for the C-H sigma orbitals in ethylene). GGAs improve on LDA by predicting physically-reasonable (negative) genuine exchange energies for all but inner core orbitals; the only exception we found is the $\sigma$-bonding orbital in H$_2$ (note that H$_2$ was excluded from \cref{table:genex_errors} since its exact and HF genuine exchange energies vanish, but it's included in the molecular dataset used in the rest of the manuscript).

\begin{table*}
    \centering
    \begin{tabular}{cccccccccc}
    \toprule
    \toprule
    C$_2$H$_4$ & {} & \multicolumn{4}{c}{Gross Exchange} & \multicolumn{4}{c}{Genuine Exchange} \\ \cmidrule(lr){3-6} \cmidrule(lr){7-10} 
    orbital & $E_{\mathrm{SI}}$& HFx & LDAx & B88 & PBEx & HFx & LDAx & B88 & PBEx \\
    \hline
    1s & 3.566 & -3.599 & -3.123 & -3.523 & -3.504 & -0.033 & 0.443 & 0.043 & 0.062 \\ 
    1s & 3.566 & -3.599 & -3.123 & -3.523 & -3.504 & -0.033 & 0.443 & 0.043 & 0.062 \\ 
    sigma & 0.705 & -0.762 & -0.701 & -0.772 & -0.769 & -0.057 & 0.005 & -0.067 & -0.062 \\ 
    sigma & 0.705 & -0.762 & -0.701 & -0.772 & -0.769 & -0.057 & 0.005 & -0.067 & -0.062 \\ 
    sigma & 0.705 & -0.762 & -0.701 & -0.772 & -0.769 & -0.057 & 0.005 & -0.067 & -0.062 \\ 
    sigma & 0.705 & -0.762 & -0.701 & -0.772 & -0.769 & -0.057 & 0.005 & -0.067 & -0.062 \\ 
    banana & 0.624 & -0.749 & -0.738 & -0.802 & -0.797 & -0.125 & -0.113 & -0.177 & -0.172 \\ 
    banana & 0.624 & -0.749 & -0.738 & -0.802 & -0.797 & -0.125 & -0.113 & -0.177 & -0.172 \\ 
    \hline
    total: & 11.202 & -11.745 & -10.524 & -11.740 & -11.669 & -0.543 & 0.678 & -0.538 & -0.467 \\ 
    \multicolumn{10}{c}{} \\
    CO & {} & \multicolumn{4}{c}{Gross Exchange} & \multicolumn{4}{c}{Genuine Exchange} \\ \cmidrule(lr){3-6} \cmidrule(lr){7-10} 
    orbital & $E_{\mathrm{SI}}$& HFx & LDAx & B88 & PBEx & HFx & LDAx & B88 & PBEx \\
    \hline
    O 1s & 4.857 & -4.927 & -4.290 & -4.800 & -4.774 & -0.070 & 0.567 & 0.058 & 0.083 \\ 
    C 1s & 3.586 & -3.614 & -3.135 & -3.543 & -3.523 & -0.028 & 0.451 & 0.042 & 0.062 \\ 
    O lone pair & 0.900 & -1.081 & -1.016 & -1.113 & -1.106 & -0.182 & -0.116 & -0.214 & -0.206 \\ 
    banana & 0.792 & -0.993 & -0.975 & -1.059 & -1.053 & -0.202 & -0.184 & -0.268 & -0.261 \\ 
    banana & 0.792 & -0.993 & -0.975 & -1.059 & -1.053 & -0.202 & -0.184 & -0.268 & -0.262 \\ 
    banana & 0.792 & -0.993 & -0.975 & -1.059 & -1.053 & -0.202 & -0.184 & -0.268 & -0.261 \\ 
    C lone pair & 0.648 & -0.703 & -0.651 & -0.733 & -0.726 & -0.055 & -0.003 & -0.085 & -0.078 \\ 
    \hline
    total: & 12.365 & -13.305 & -12.017 & -13.367 & -13.289 & -0.940 & 0.348 & -1.002 & -0.924 \\ 
    \multicolumn{10}{c}{} \\
    HF & {} & \multicolumn{4}{c}{Gross Exchange} & \multicolumn{4}{c}{Genuine Exchange} \\ \cmidrule(lr){3-6} \cmidrule(lr){7-10} 
    orbital & $E_{\mathrm{SI}}$& HFx & LDAx & B88 & PBEx & HFx & LDAx & B88 & PBEx \\
    \hline
    F 1s & 5.494 & -5.592 & -4.877 & -5.433 & -5.405 & -0.098 & 0.617 & 0.061 & 0.090 \\ 
    lone pair & 1.018 & -1.240 & -1.181 & -1.291 & -1.282 & -0.222 & -0.163 & -0.273 & -0.264 \\ 
    lone pair & 1.018 & -1.240 & -1.181 & -1.291 & -1.282 & -0.222 & -0.163 & -0.273 & -0.264 \\ 
    lone pair & 1.018 & -1.240 & -1.181 & -1.291 & -1.282 & -0.222 & -0.163 & -0.273 & -0.264 \\ 
    sigma & 0.927 & -1.117 & -1.069 & -1.165 & -1.157 & -0.190 & -0.142 & -0.238 & -0.230 \\ 
    \hline
    total: & 9.475 & -10.43 & -9.490 & -10.471 & -10.409 & -0.954 & -0.014 & -0.996 & -0.934 \\ 

    \bottomrule
    \bottomrule
    \end{tabular}
    \caption{Breakdown of exchange energy by orbital for three representative molecules containing single, double, and triple covalent bonds. All energies in $E_{\mathrm{h}}$.}
    \label{table:molecule_per-orbital}
\end{table*}

The associated errors in the per-orbital genuine exchange for a larger set of LDA, GGAs, and meta-GGAs (\cref{table:molecules_per-orbital-error_ks}) illustrate trends similar to those found in atoms. Whereas LDA does not benefit from core-valence cancellation of genuine exchange energies, GGAs (conventional and meta-) both benefit tremendously from the cancellation. Although meta-GGAs per-orbital errors are substatially smaller than those of GGAs, the total errors of meta-GGAs are not always smaller than the GGA counterparts. Similarly, the M06-Lx meta-GGA benefits from error cancellation more than the other meta-GGAs, despite its per-orbital errors not being systematically smaller.

\begin{table*}
    \centering
    \begin{tabular}{lrrrrrr}
    \toprule
    \toprule
    C$_2$H$_4$ & \multicolumn{6}{c}{} \\
    orbital & LDAx & B88 & PBEx & revTPSSx & revSCANx & M06-Lx \\
    \hline
    1s (C) & 0.476 & 0.076 & 0.095 & 0.025 & 0.018 & 0.039 \\ 
    1s (C) & 0.476 & 0.076 & 0.095 & 0.025 & 0.018 & 0.039 \\ 
    sigma (C-H) & 0.061 & -0.010 & -0.005 & -0.002 & -0.003 & -0.004 \\ 
    sigma (C-H) & 0.061 & -0.010 & -0.005 & -0.002 & -0.003 & -0.004 \\ 
    sigma (C-H) & 0.061 & -0.010 & -0.005 & -0.002 & -0.003 & -0.004 \\ 
    sigma (C-H) & 0.061 & -0.010 & -0.005 & -0.002 & -0.003 & -0.004 \\ 
    banana (C-C) & 0.011 & -0.052 & -0.047 & -0.031 & -0.023 & -0.024 \\ 
    banana (C-C) & 0.011 & -0.052 & -0.047 & -0.031 & -0.023 & -0.024 \\ 
    \hline
    total & 1.221 & 0.005 & 0.076 & -0.021 & -0.022 & 0.014 \\ 
    \multicolumn{7}{c}{} \\
    CO & \multicolumn{6}{c}{} \\
    orbital & LDAx & B88 & PBEx & revTPSSx & revSCANx & M06-Lx \\
    \hline
    1s (O) & 0.637 & 0.127 & 0.153 & 0.050 & 0.033 & 0.069 \\ 
    1s (C) & 0.479 & 0.070 & 0.090 & 0.020 & 0.015 & 0.035 \\ 
    banana (C-O) & 0.018 & -0.066 & -0.060 & -0.040 & -0.036 & -0.034 \\ 
    banana (C-O) & 0.018 & -0.066 & -0.060 & -0.040 & -0.036 & -0.034 \\ 
    banana (C-O) & 0.018 & -0.066 & -0.060 & -0.040 & -0.036 & -0.034 \\ 
    lone pair (O) & 0.065 & -0.032 & -0.025 & -0.009 & -0.011 & -0.006 \\ 
    lone pair (C) & 0.052 & -0.030 & -0.023 & -0.013 & -0.001 & -0.004 \\ 
    \hline
    total & 1.288 & -0.062 & 0.016 & -0.072 & -0.071 & -0.007 \\ 
    \multicolumn{7}{c}{} \\
    HF & \multicolumn{6}{c}{} \\
    orbital & LDAx & B88 & PBEx & revTPSSx & revSCANx & M06-Lx \\
    \hline
    1s (F) & 0.715 & 0.159 & 0.187 & 0.068 & 0.043 & 0.088 \\ 
    sigma (F-H) & 0.048 & -0.048 & -0.040 & -0.023 & -0.026 & -0.025 \\ 
    lone pair (F) & 0.059 & -0.051 & -0.042 & -0.023 & -0.025 & -0.022 \\ 
    lone pair (F) & 0.059 & -0.051 & -0.042 & -0.023 & -0.025 & -0.022 \\ 
    lone pair (F) & 0.059 & -0.051 & -0.042 & -0.023 & -0.025 & -0.022 \\ 
    \hline
    total & 0.940 & -0.042 & 0.020 & -0.024 & -0.058 & -0.003 \\ 
    \bottomrule
    \bottomrule
    \end{tabular}
    \caption{Breakdown of errors in genuine exchange energy by orbital using KS for some molecules from G2 set. All energies in $E_{\mathrm{h}}$.}
    \label{table:molecules_per-orbital-error_ks}
\end{table*}

\begin{figure*}
    \centering
    \includegraphics[width=0.9\textwidth]{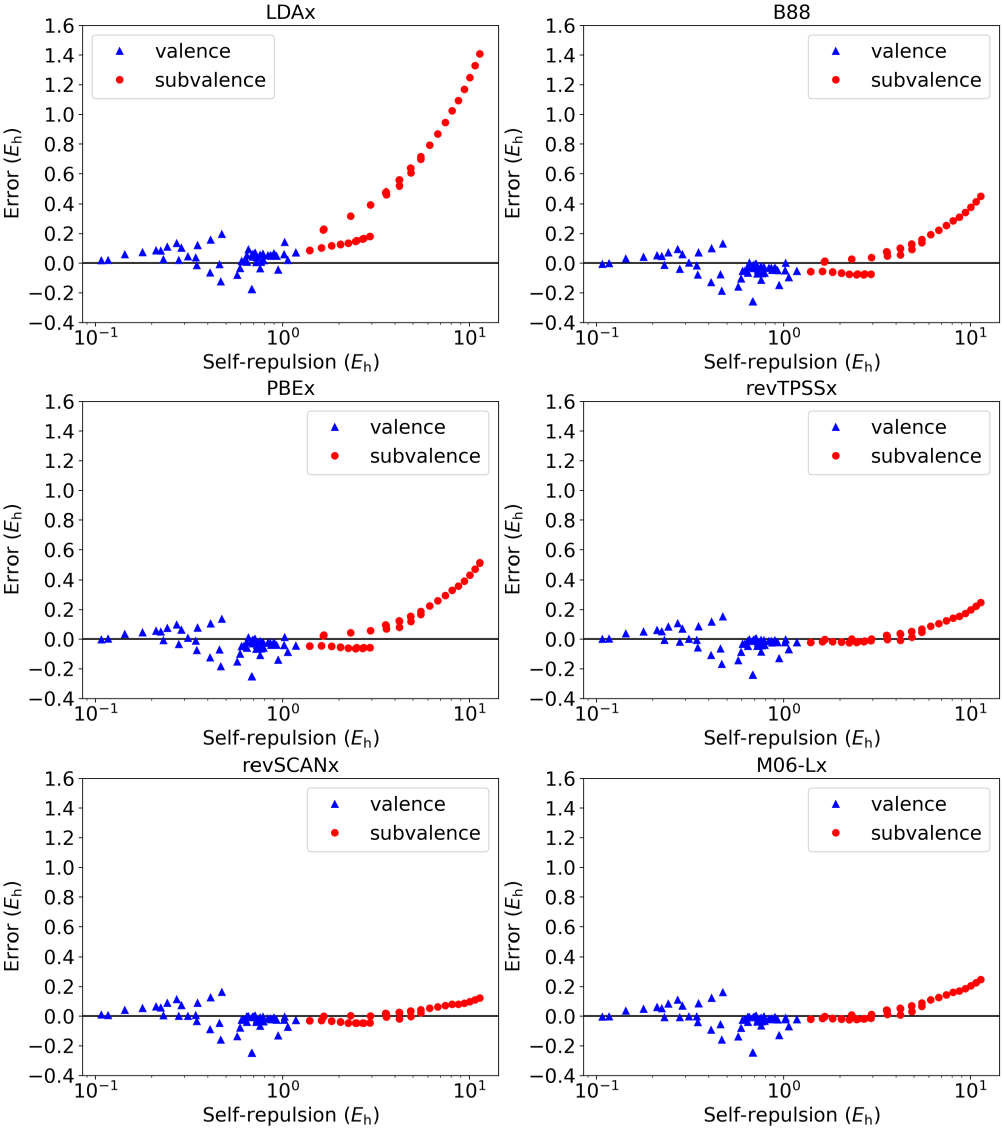}
    \caption{Orbital exchange energy error vs self-repulsion energy for the combined testset of atoms and molecules using select LDA, GGA, and meta-GGA functionals.}
    \label{fig:ks_error_scatter}
\end{figure*}

A useful insight is provided by plotting $\Delta [\bar{E}_\mathrm{x}^\mathrm{KS}]_i$ against the per-orbital self-repulsion, $\braket{ii}{ii}/2$. \cref{fig:ks_error_scatter} contains such plots aggregating the data from all atomic and molecular computations, for an illustrative set of functionals. The orbitals are further broken down into ``valence'' and ``subvalence'' orbitals (the latter including $n=2$ orbitals in atoms Na-Ar). Several trends are easily identified. Whereas the vast majority of LDA errors are positive (i.e., the exchange-induced stabilitization is underestimated by LDA), the errors are smaller for GGAs and especially meta-GGAs, and benefit from cancellation of positive errors for core orbitals (with self-repulsion energy greater than $3 E_\mathrm{h}$) and negative errors for valence and subvalence orbitals (with self-repulsion energy between 0.5 and 3 $E_\mathrm{h}$). The errors for orbitals with very low self-repulsion energies (less than 0.5 $E_\mathrm{h}$) are not systematic.

\subsection{Orbital Breakdown of PZ-Corrected KS Exchange Energies} 

Representative snapshots of errors in genuine exchange energies obtained with PZ-corrected KS functionals (LDA, GGAs, and meta-GGAs) are presented in \cref{table:atom_per-orbital-error_pz} and \cref{table:molecules_per-orbital-error_pz}, respectively. The rest of the data is available in the Supplementary Information.

\begin{figure}
    \centering
    \includegraphics[width=0.9\columnwidth]{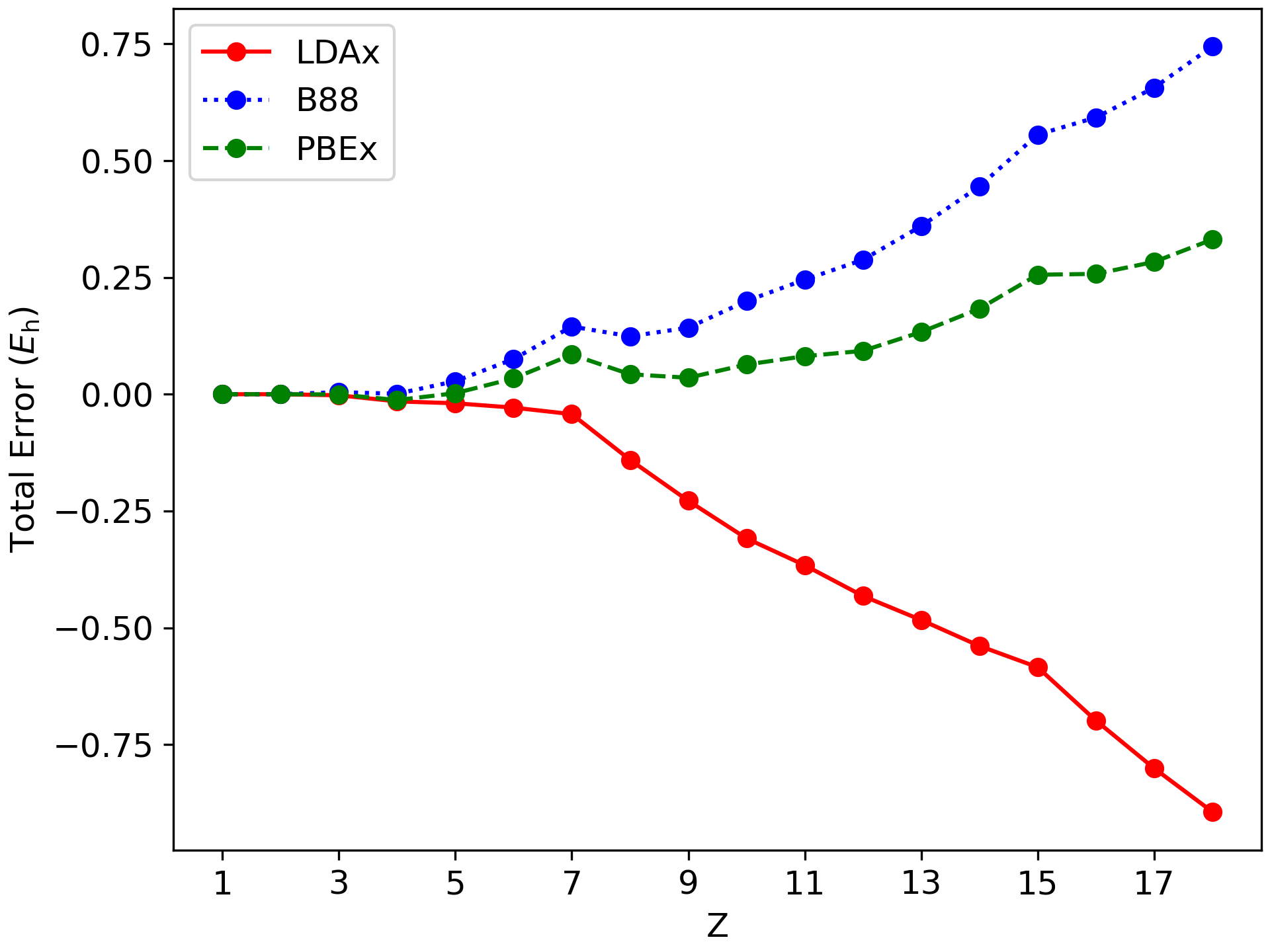}
    \caption{Errors of PZ-corrected KS genuine exchange energies for atoms $Z \leq 18$.} 
    \label{fig:atoms_total_ex_error_pz}
\end{figure}

\begin{figure}
    \centering
    \includegraphics[width=0.9\columnwidth]{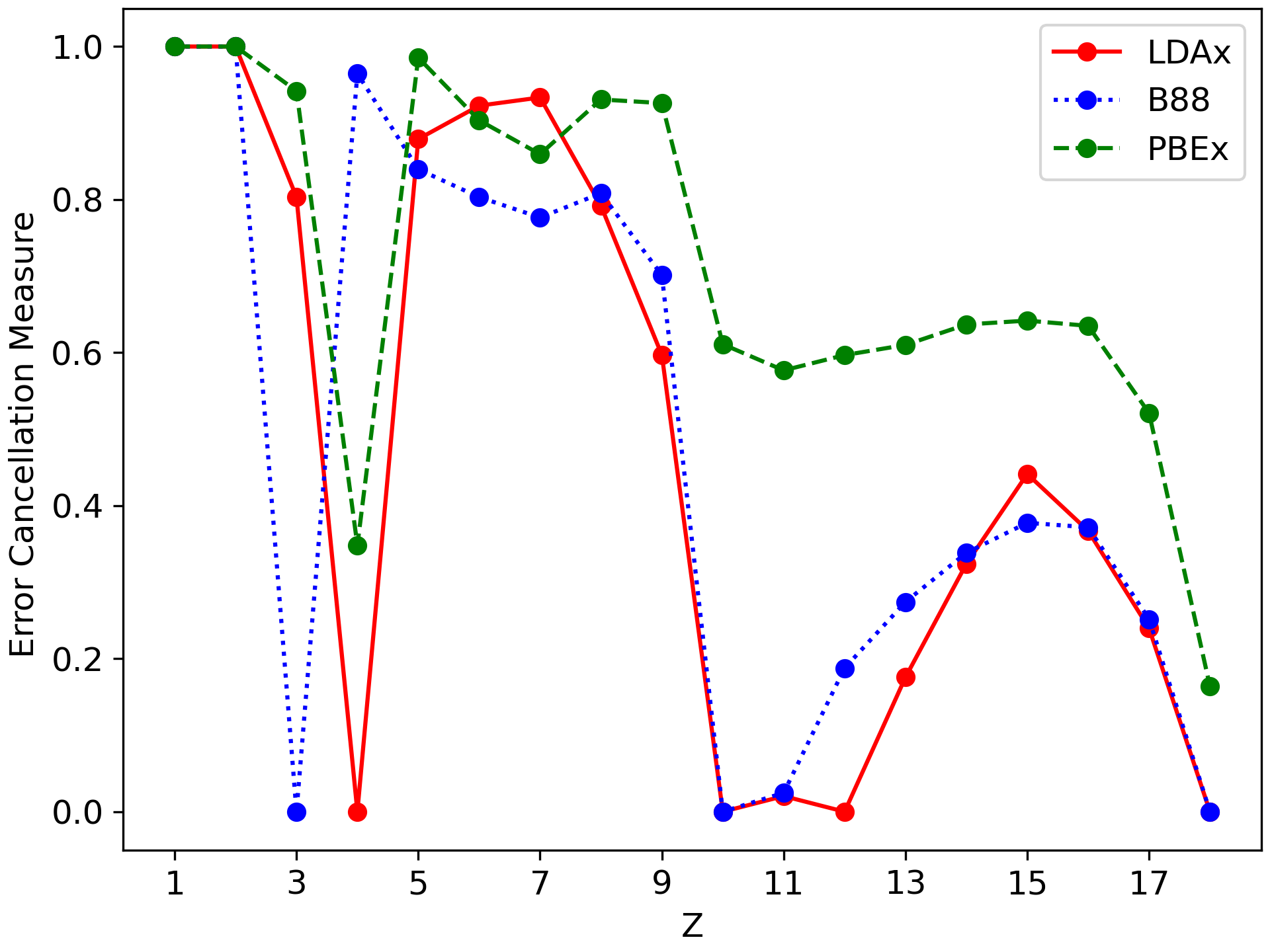}
    \caption{The error cancellation measure (\cref{eq:err-cancellatior-ratio}) for PZ-corrected KS genuine exchange energies of atoms $Z 
    \leq 18$.} 
    \label{fig:atoms_error_cancel_pz}
\end{figure}

As is known, the PZ correction greatly improves the accuracy of LDAx, but actually makes the GGAs worse\cite{VRG:vydrov:2004:JCP}.
Indeed, whereas LDAx systematically (and severely) underestimated the genuine exchange energies (see \cref{fig:atoms_total_ex_error}), PZ-LDAx systematically overestimates them, but to a lesser degree (see \cref{fig:atoms_total_ex_error_pz}).
In fact, unlike LDAx, PZ-LDAx exhibits some error cancellation (compare \cref{fig:atoms_error_cancel} to \cref{fig:atoms_error_cancel_pz}).
The effect of PZ correction for GGAs is uneven: whereas the errors are significantly reduced for valence orbitals, they are mostly unchanged for the core orbitals. For the outer core ($n=2$) orbitals the PZ correction ``overcorrects'' the KS genuine exchange errors. The end result is that the systematic cancellation of errors between core and valence observed with GGAs no longer holds with PZ-corrected GGAs. The effect of PZ correction for meta-GGAs is less systematic, and strongly depends on the functional. E.g., PZ correction makes M06-Lx significantly less accurate, remarkably by increasing significantly the errors for core orbitals and completely spoiling the errors cancellations that were observed for non-PZ M06-Lx. 

\begin{figure*}
    \centering
    \includegraphics[width=0.9\textwidth]{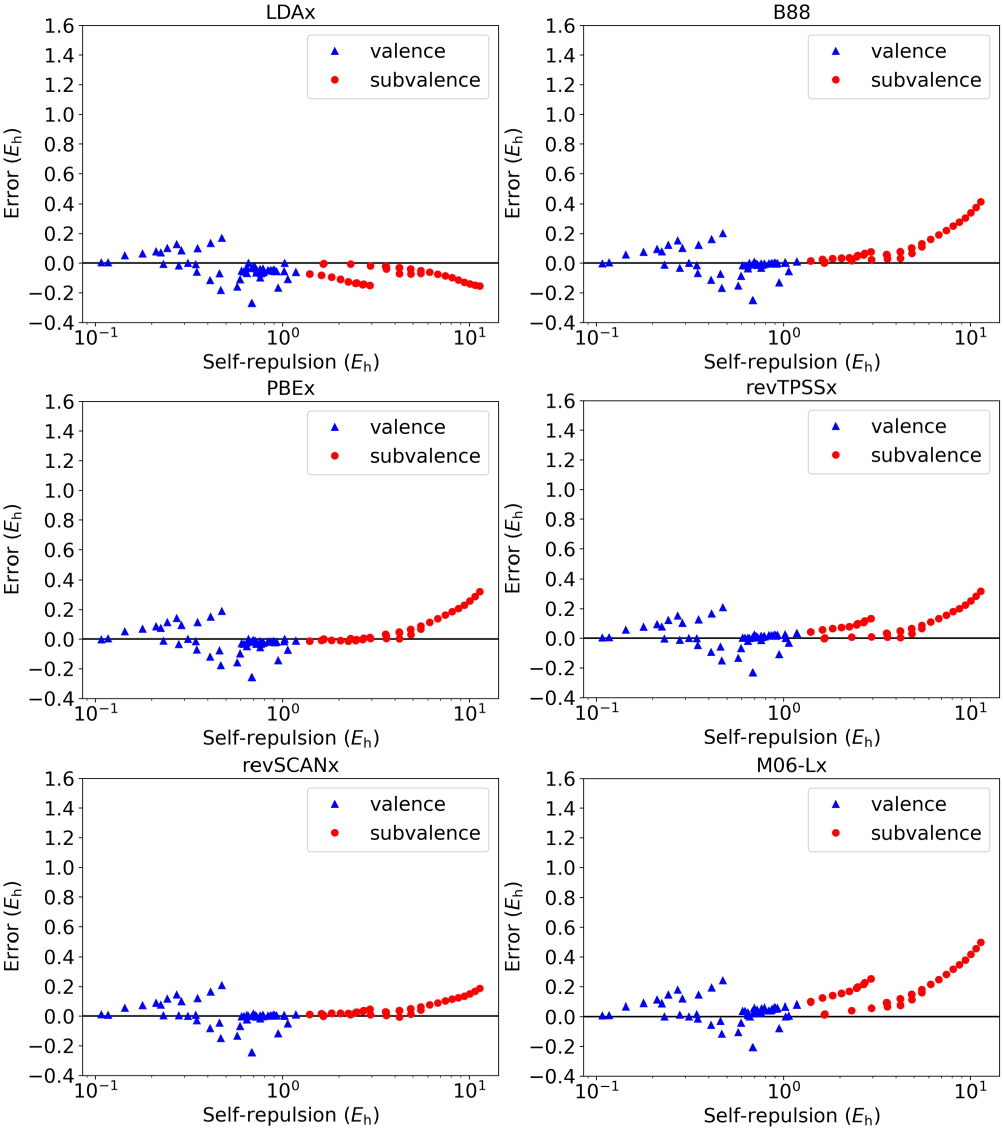}
    \caption{Orbital exchange energy error vs self-repulsion energy for the combined set of atoms and molecules using select PZ-corrected LDA, GGA, and meta-GGA functionals.}
    \label{fig:pz_error_scatter}
\end{figure*}

Plots of $\Delta [\bar{E}_\mathrm{x}^\mathrm{PZ}]_i$ against the per-orbital self-repulsion are again helpful (\cref{fig:pz_error_scatter}) by emphasizing that the PZ correction (except for LDA) does not significantly improve the genuine exchange description for the inner core orbitals. Interestingly, PZ-LDAx overestimates the genuine exchange for the inner core orbitals, but underestimates them otherwise (just like the standard KS functionals). Clearly, the performance of PZ corrections strongly depends on the functional, especially for the meta-GGAs: PZ-revSCANx genuine exchange of valence and outer core orbitals is remarkably accurate, whereas the accuracy of PZ-M06-Lx and PZ-revTPSSx is rather poor.

Although the orbital decomposition reveals a strickingly-uneven effects of PZ corrections, the finer details of performance of PZ-corrected KS for different types of orbitals may open a path to further improvements.

\begin{table*}
    \centering
    \begin{tabular}{lrrrrrr}
    \toprule
    \toprule
    Ne & \multicolumn{6}{c}{} \\
    orbital & LDAx & B88 & PBEx & revTPSSx & revSCANx & M06-Lx \\
    \hline
    1s & -0.061 & 0.161 & 0.114 & 0.109 & 0.076 & 0.215 \\ 
    2sp$^3$ & -0.062 & 0.010 & -0.012 & 0.033 & 0.010 & 0.080 \\ 
    2sp$^3$ & -0.062 & 0.010 & -0.012 & 0.033 & 0.010 & 0.080 \\ 
    2sp$^3$ & -0.062 & 0.010 & -0.012 & 0.033 & 0.010 & 0.080 \\ 
    2sp$^3$ & -0.062 & 0.010 & -0.012 & 0.033 & 0.010 & 0.080 \\ 
    \hline
    total & -0.309 & 0.200 & 0.064 & 0.243 & 0.116 & 0.536 \\ 
    \multicolumn{7}{c}{} \\
    Ar & \multicolumn{6}{c}{} \\
    orbital & LDAx & B88 & PBEx & revTPSSx & revSCANx & M06-Lx \\
    \hline
    1s & -0.155 & 0.414 & 0.320 & 0.317 & 0.187 & 0.498 \\ 
    2sp$^3$ & -0.151 & 0.076 & 0.011 & 0.132 & 0.046 & 0.251 \\ 
    2sp$^3$ & -0.151 & 0.076 & 0.011 & 0.132 & 0.046 & 0.251 \\ 
    2sp$^3$ & -0.151 & 0.076 & 0.011 & 0.132 & 0.046 & 0.251 \\ 
    2sp$^3$ & -0.151 & 0.076 & 0.011 & 0.132 & 0.046 & 0.251 \\ 
    3sp$^3$ & -0.034 & 0.007 & -0.008 & 0.023 & 0.015 & 0.057 \\ 
    3sp$^3$ & -0.034 & 0.007 & -0.008 & 0.023 & 0.015 & 0.057 \\ 
    3sp$^3$ & -0.034 & 0.007 & -0.008 & 0.023 & 0.015 & 0.057 \\ 
    3sp$^3$ & -0.034 & 0.007 & -0.008 & 0.023 & 0.015 & 0.057 \\ 
    
    \hline
    total & -0.894 & 0.744 & 0.331 & 0.935 & 0.430 & 1.730 \\ 
    \bottomrule
    \bottomrule
    \end{tabular}
    \caption{Per-orbital errors in PZ-corrected KS genuine exchange energies for atoms Ne and Ar. All energies in $E_{\mathrm{h}}$.}
    \label{table:atom_per-orbital-error_pz}
\end{table*}

\begin{table*}[hbp]
    \centering
    \begin{tabular}{lrrrrrr}
    \toprule
    \toprule
    C$_2$H$_4$ & \multicolumn{6}{c}{} \\
    orbital & LDAx & B88 & PBEx & revTPSSx & revSCANx & M06-Lx \\
    \hline
    1s (C) & -0.023 & 0.057 & 0.031 & 0.034 & 0.028 & 0.092 \\ 
    1s (C) & -0.023 & 0.057 & 0.031 & 0.034 & 0.028 & 0.092 \\ 
    sigma (C-H) & -0.034 & -0.006 & -0.015 & 0.009 & 0.002 & 0.025 \\ 
    sigma (C-H) & -0.034 & -0.006 & -0.015 & 0.009 & 0.002 & 0.025 \\ 
    sigma (C-H) & -0.034 & -0.006 & -0.015 & 0.009 & 0.002 & 0.025 \\ 
    sigma (C-H) & -0.034 & -0.006 & -0.015 & 0.009 & 0.002 & 0.025 \\ 
    banana (C-C) & -0.057 & -0.018 & -0.032 & 0.002 & -0.002 & 0.037 \\
    banana (C-C) & -0.057 & -0.018 & -0.032 & 0.002 & -0.002 & 0.037 \\ 
    \hline
    total & -0.295 & 0.054 & -0.063 & 0.109 & 0.061 & 0.355 \\ 
    \multicolumn{7}{c}{} \\
    CO & \multicolumn{6}{c}{} \\
    orbital & LDAx & B88 & PBEx & revTPSSx & revSCANx & M06-Lx \\
    \hline
    1s (O) & -0.041 & 0.101 & 0.065 & 0.065 & 0.049 & 0.147 \\ 
    1s (C) & -0.022 & 0.051 & 0.026 & 0.030 & 0.026 & 0.090 \\ 
    banana (C-O) & -0.062 & -0.007 & -0.026 & 0.011 & -0.001 & 0.059 \\ 
    banana (C-O) & -0.062 & -0.007 & -0.026 & 0.011 & -0.001 & 0.059 \\ 
    banana (C-O) & -0.062 & -0.007 & -0.026 & 0.011 & -0.001 & 0.059 \\ 
    lone pair (O) & -0.044 & 0.002 & -0.014 & 0.023 & 0.009 & 0.060 \\ 
    lone pair (C) & -0.032 & -0.017 & -0.025 & 0.002 & 0.007 & 0.029 \\ 
    \hline
    total & -0.324 & 0.117 & -0.026 & 0.154 & 0.087 & 0.504 \\ 
    \multicolumn{7}{c}{} \\
    HF & \multicolumn{6}{c}{} \\
    orbital & LDAx & B88 & PBEx & revTPSSx & revSCANx & M06-Lx \\
    \hline
    1s (F) & -0.051 & 0.130 & 0.088 & 0.086 & 0.062 & 0.180 \\ 
    sigma (F-H) & -0.056 & -0.001 & -0.018 & 0.022 & 0.004 & 0.056 \\ 
    lone pair (F) & -0.056 & 0.003 & -0.016 & 0.024 & 0.006 & 0.065 \\ 
    lone pair (F) & -0.056 & 0.003 & -0.016 & 0.024 & 0.006 & 0.065 \\ 
    lone pair (F) & -0.056 & 0.003 & -0.016 & 0.024 & 0.006 & 0.065 \\ 
    \hline
    total & -0.276 & 0.138 & 0.023 & 0.181 & 0.084 & 0.431 \\ 
    \bottomrule
    \bottomrule
    \end{tabular}
    \caption{Per-orbital errors in PZ-corrected KS genuine exchange energies for three representative molecules containing single, double, and triple covalent bonds. All energies in $E_{\mathrm{h}}$.}
    \label{table:molecules_per-orbital-error_pz}
\end{table*}

\section{Summary and Outlook}
\label{section:summary}

To obtain a detailed quantitative portrait of self-interaction errors in KS and other mean-field models of electronic structure we proposed to identify the exact genuine (self-interaction-free) mean-field exchange energy as the difference between the energies of Hartree-Fock and orthogonal Hartree models. The difference between the gross HF exchange energy and the genuine exchange thereby uniquely defines the self-interaction component of the exchange energy for any many-electron system. By showing numerically that the differences between orbital subspaces of HF and Hartree models are small, and the Edmiston-Ruedenberg-localized HF orbitals approximate the exact Hartree orbitals most closely (undoubtedly due to their strong formal connection), we devised a decomposition of the genuine exchange energies of mean-field models (and other energy components) into their orbital contributions. Initial numerical investigation on light atoms and small molecules revealed a plethora of insights:
\begin{itemize}
\item {\bf LDA} severely underestimates genuine exchange effects for both valence and core orbitals. Perdew-Zunger self-interaction correction overcorrects the deficiencies of the LDA for both valence and core regions.
\item {\bf GGA} functionals underestimate genuine exchange effects for the core orbitals, but nearly perfectly compensate by overestimating them for valence (and outer core) orbitals. The remarkable cancellation of errors produces total exchange energies far more accurate than the exchange energies for individual orbitals. Perdew-Zunger self-interaction correction addresses well the deficiencies of GGAs for valence orbitals, but not for the subvalence orbitals.
\item {\bf Meta-GGA} functionals improve on GGAs by reducing errors in genuine exchange, but their valence/core traits are unchanged. The errors cancellation is less systematic, and the effect of PZ correction often increases the errors in genuine exchange. Performance of PZ-corrected meta-GGAs also strongly depends on the functional.
\item {\bf Hybrid} functionals reduce errors in genuine exchange, by simple reduction of the semilocal contribution to the functional. But due to the general-purpose hybrid functionals usually including a modest fraction ($<30\%$) of exact exchange, such reductions will be equaly modest, and will not affect the cancellation of errors in genuine exchange between core and valence regions.
\end{itemize}



Although the proposed orbital decomposition reveals hidden flaws in the performance of popular KS functionals and their PZ-corrected counterparts, a more detailed performance profile may also open a path to further improvements. For example, the exact 2-electron condition for the per-orbital genuine exchange energies (\cref{eq:Ebar-2e}) could be used as a guide to improved functionals in the future. Some work along these direction has already started in our laboratory.

\section*{Supplementary Information}

Breakdown of the exchange energy by orbital for both KS and PZ methods for a variety of functionals (for atoms Z $\le$ 18 and molecules) is provided as a plain text file.


\section*{Acknowledgments}

This work was supported by the U.S. Department of Energy via award DE-SC0022327.

We would like to acknoweledge helpful in-depth discussions with Victor Staroverov during the early stages of this work.


\appendix
\section{The self-consistent solver for the orthogonal Hartree method.}
\label{section:solver-hartree}

The original method of Hartree\cite{VRG:hartree:1928:MPCPS,VRG:hartree:1928:MPCPSa} produced nonorthogonal orbitals. To satisfy the noninteracting fermionic $N$-representability conditions it is desirable to constrain the orbitals entering the Hartree product to be \textit{orthonormal}.
The orthogonal Hartree method is distinct from the original method of Hartree but usually produces results quite close to the latter.
It has been investigated but only briefly.\cite{VRG:christiansen:1977:JCP,VRG:harris:1978:IJoQC,VRG:perera:1994:IJoQC}
Our objective is to compute the Hartree product,
\begin{align}
\Phi^\mathrm{H}(\mathbf{x}_1, \mathbf{x}_2, ... \mathbf{x}_N) = \phi_1(\mathbf{x}_1) \phi_2(\mathbf{x}_2) ... \phi_N(\mathbf{x}_N),
\end{align}
that minimizes \cref{eq:EH} subject to the constraint that $\braket{i}{j} = \delta_i^j = \begin{cases} 1, i=j \\ 0, i \neq j \end{cases}.$

Although methods exist which express the Hartree orbital optimization problem as a self-consistent (nonlinear) eigenvalue problem,\cite{VRG:harris:1978:IJoQC} here we used a solver that directly  minimizes the Hartree energy (\cref{eq:EH}) by unitary rotations. From the perspective of manifold optimization, the solution space of the orthogonal Hartree method is a Stiefel manifold, unlike the Grassman manifold of the SCF; however, the same basic techniques can be applied to nonlinear optimization in both contexts. In the context of conventional SCF such solvers have evolved from relatively simple\cite{VRG:wong:1995:JCC} to quite elaborate (e.g., see recent work\cite{VRG:helmich-paris:2021:JCP,VRG:slattery:2024:PCCP,VRG:sethio:2024:JPCA}). The Hartree solver deployed here is a simple nonlinear approximate conjugate gradient descent solver using approximate line-search.

$N$ Hartree orbitals and their $M-N$ orthogonal complements are expanded in terms of $M$ linearly-independent AOs (linear dependencies can be handled straightforwardly as a pre-processing step). The target AO coefficients are the columns of a square matrix, $\mathbf{C} \in \mathbb{R}^{M\times M}$, that can be expressed as a $\mathbb{R}^{M\times M} $ unitary applied to the starting set of orthonormal orbitals, $\mathbf{C^{(0)}}$. With standard {\em real} exponential parametrization of the unitary in terms of a square (upper or lower) triangular $\bm{\kappa} \in \mathbb{R}^{M \times M}$
\begin{align}
\label{eq:Cexp}
\mathbf{C}=\mathbf{C^{(0)}} \exp(\bm{\kappa} - \bm{\kappa}^{\dagger}),
\end{align}
Plugging \cref{eq:Cexp} into the Hartree energy expression \cref{eq:EH} and differentiating with respect to matrix elements of $\bm{\kappa}$ yields 2 types of nonzeros. The first is due to orbital rotations involving an occupied orbital $i$ and an unoccupied orbital $a$:
\begin{align}
\frac{\partial E^\mathrm{H}}{\bm{\kappa}_a^i} = 2 \left ( \langle a | \hat{h} | i \rangle + \sum_{j\neq i} \langle aj|ij \rangle \right )
\end{align}
The second is due to the orbital rotations involving two occupied orbitals, $i$ and $j$:
\begin{align}
\frac{\partial E_\mathrm{H}}{\bm{\kappa}_j^i} = 2 \left ( \langle jj | ij \rangle - \langle ii | ji \rangle \right ).
\end{align}
In conventional SCF the occ-occ rotations do not change the energy since a Slater determinant is invariant with respect to unitary transformation of its columns. The Hartree product is not invariant with respect to such rotations, hence the nonzero occ-occ gradient. 
As has been noted by others,\cite{VRG:perera:1994:IJoQC} stationarity of the Hartree energy with respect to occ-occ rotations is equivalent to the stationarity condition of the the Edmiston-Ruedenberg (ER) localization functional\cite{VRG:edmiston:1963:RMP}.
The connection is not coincidental since the ER localization was designed to maximize $J^\mathrm{SI}$! For a fixed span of occupied orbitals (hence fixed 1-electron energy and the gross Coulomb energy $J$) maximization of $J^\mathrm{SI}$ is equivalent to minimization of Hartree energy \cref{eq:EH}, as seen clearly from \cref{eq:JH}.
However, since the occ-uocc gradient of the Hartree energy differs from that of the Hartree-Fock energy (due to the absence of the genuine exchange effects in the former), the span of Hartree orbitals will in general differ from the span of Hartree-Fock orbitals.
However, for well-behaved systems the differences between spans of Hartree and HF orbitals are minor.
Thus the Hartree orbitals will be localized and similar to the ER Hartree-Fock orbitals.
Due to this connection, for non-self-consistent calculations, we will use ER localized HF orbitals as sufficiently good approximations to the true Hartree orbitals.
In \cref{subsection:results-hartree} we demonstrate the performance of this approximation.

\bibliography{sas-refs.bib,maybe_thesis,vrg_refs}

\end{document}